%% file: main.tex
\documentclass[11pt]{article}
\usepackage[T1]{fontenc}
\usepackage[utf8]{inputenc}
\usepackage{mathptmx}
\usepackage[margin=1.1in]{geometry}
\usepackage{amsmath,amssymb}
\usepackage{graphicx}
\usepackage{booktabs}
\usepackage{caption}
\usepackage{subcaption}
\usepackage{setspace}
\usepackage{microtype}
\usepackage{enumitem}
\usepackage{xcolor}
\usepackage[round,authoryear]{natbib}
\usepackage[colorlinks=true,linkcolor=blue!60!black,citecolor=blue!60!black,urlcolor=blue!60!black]{hyperref}
\usepackage{appendix}
\usepackage[section]{placeins}   
\graphicspath{{figures/}}
\input{numbers.tex}

\setlist{nosep}
\newcommand{\rev}[1]{reviewer~#1}

\title{\Large\textbf{Reliability of AI Agents:\\ Rater Effects, Drift, and the Return to an Evaluation Program}}
\author{Liu Zhang \ \ \   Mark Esposito }
\date{\normalsize Working paper, October 2026. Preliminary; comments welcome.}

\begin{document}
\maketitle
\begin{abstract}
\noindent Firms increasingly evaluate deployed AI agents with repeated human ratings, but score changes can reflect the measurement process as well as the agent, and firms rarely know how long an evaluation remains informative. We study reliability as a latent state using \nProd{} rubric-scored production interviews from a deployed voice-and-video interviewing agent, overlapping human reviewers, the agent's change log, and support tickets. Reviewer severity was large: two reviewers evaluating the same batches differed by \feRevBAbs{} standard deviations on the composite index, and a change in reviewer composition flattened the raw trend. After adjusting for reviewer effects and smoothing batch-level estimates, evaluated reliability rose by \latentIndexChange{} standard deviations from March to August 2026. Reliability also moved materially between logged deploys. At the estimated rate of unaccounted movement, forecast uncertainty reaches the size of the largest observed deploy contrast after about five weeks, although this horizon is imprecisely estimated and reflects a small number of step-like changes. An independent business outcome measure moved in the same direction: interview-related support tickets fell about \tickFullAbs\% per month relative to technical-issue tickets. These data do not identify the causal effect of evaluation on performance. They instead show that longitudinal AI evaluation requires three checks: score comparability across raters, evidence currency under drift, and corroboration with consequential operational outcomes.

\noindent\textbf{Keywords:} AI agents, evaluation, reliability, rater effects, latent variables, drift, measurement, AI interview.\\
\noindent\textbf{JEL:} C38, D83, L86, M15, O33.
\end{abstract}
\newpage

\newpage

\input{sections/01_intro}
\input{sections/02_framework}
\input{sections/03_setting}
\input{sections/04_improving}
\input{sections/05_drift}
\input{sections/06_decisions}

\input{sections/07_conclusion}

\clearpage
\bibliographystyle{plainnat}
\bibliography{references}

\clearpage
\begin{appendices}
\input{sections/appendix}

\end{appendices}
\end{document}

%% file: numbers.tex
\newcommand{\nStabAll}{3,521}
\newcommand{\nProd}{2,611}
\newcommand{\nBatches}{14}

\newcommand{\nStaging}{910}

\newcommand{\feRevBAbs}{0.79}
\newcommand{\feRevBSe}{0.05}

\newcommand{\feRevBCrit}{\ensuremath{-}2.2}

\newcommand{\rTwoBatch}{1.4}

\newcommand{\rTwoBatchTrainer}{16.2}

\newcommand{\gapPooled}{\ensuremath{-}0.77}

\newcommand{\gapPooledSe}{0.03}

\newcommand{\gapQ}{19.0}

\newcommand{\gapQp}{0.01}
\newcommand{\gapN}{9}
\newcommand{\gapMin}{\ensuremath{-}0.90}

\newcommand{\gapMax}{\ensuremath{-}0.53}

\newcommand{\sevLatencyRevB}{\ensuremath{-}1.40}

\newcommand{\sevStructureRevB}{\ensuremath{-}0.13}

\newcommand{\critRevA}{5.0}

\newcommand{\critRevB}{2.0}

\newcommand{\trendRaw}{0.002}

\newcommand{\trendRawP}{0.93}
\newcommand{\trendFe}{0.102}

\newcommand{\trendFeSe}{0.021}

\newcommand{\trendFeP}{$<$0.001}
\newcommand{\trendFeBoot}{0.002}
\newcommand{\trendRegime}{0.125}

\newcommand{\trendRegimeP}{0.002}

\newcommand{\latentIndexFirst}{0.05}

\newcommand{\latentIndexAug}{0.58}

\newcommand{\latentIndexChange}{0.53}

\newcommand{\latentIndexChangeSe}{0.07}

\newcommand{\latentIndexChangeDiag}{0.54}

\newcommand{\sdIndexWeekDiag}{0.077}

\newcommand{\offdiagCorrIndex}{0.11}

\newcommand{\latentCritFirst}{2.8}

\newcommand{\latentCritPeak}{6.3}

\newcommand{\latentCritPeakDate}{2026-05-09}
\newcommand{\latentCritAug}{2.5}

\newcommand{\obsCritLast}{0.0}

\newcommand{\obsCritPeak}{11.1}

\newcommand{\sdIndexWeek}{0.075}

\newcommand{\sdIndexWeekLo}{0.041}

\newcommand{\sdIndexWeekHi}{0.142}

\newcommand{\sdCritWeek}{0.93}

\newcommand{\sdIndexWeekFree}{0.086}

\newcommand{\sdIndexWeekFreeLo}{0.041}

\newcommand{\sdIndexWeekFreeHi}{0.180}

\newcommand{\qIndexFreeP}{$<$0.001}
\newcommand{\sdIndexWeekNoStep}{0.000}

\newcommand{\sdIndexWeekNoStepHi}{0.151}

\newcommand{\qIndexNoStepP}{0.50}
\newcommand{\sdCritWeekFree}{0.76}

\newcommand{\qCritFreeP}{0.01}
\newcommand{\sdCritWeekNoStep}{0.67}

\newcommand{\qCritNoStepP}{0.16}

\newcommand{\refreshIndexWeeksAll}{6}

\newcommand{\refreshIndexFour}{0.18}

\newcommand{\refreshIndexEight}{0.25}

\newcommand{\refreshIndexTwelve}{0.30}

\newcommand{\refreshIndexWeeks}{5}

\newcommand{\refreshCritFour}{1.9}

\newcommand{\refreshCritEight}{2.4}

\newcommand{\refreshCritTwelve}{2.9}

\newcommand{\dedStructure}{\ensuremath{-}9.3}

\newcommand{\dedStructureP}{$<$0.001}
\newcommand{\dedAvsync}{\ensuremath{-}3.6}

\newcommand{\dedAvsyncP}{0.02}
\newcommand{\dedLatency}{\ensuremath{-}3.0}

\newcommand{\dedLatencyP}{0.03}
\newcommand{\dedAudio}{\ensuremath{-}2.1}

\newcommand{\dedAudioP}{0.02}
\newcommand{\avsyncShare}{73}

\newcommand{\latencyAtMax}{55}

\newcommand{\structDedFirst}{63}

\newcommand{\structDedAug}{23}

\newcommand{\avsyncDedFirst}{16}

\newcommand{\latencyDedFirst}{28}

\newcommand{\latencyDedAug}{11}

\newcommand{\audioDedPeak}{4.8}

\newcommand{\aprIndex}{0.20}

\newcommand{\aprIndexP}{0.008}
\newcommand{\aprCrit}{4.7}

\newcommand{\aprImpr}{\ensuremath{-}0.43}

\newcommand{\aprStruct}{0.37}

\newcommand{\vendIndex}{0.02}

\newcommand{\vendImpr}{\ensuremath{-}0.05}

\newcommand{\vendCrit}{4.0}

\newcommand{\julIndex}{0.12}

\newcommand{\julStruct}{0.32}

\newcommand{\julAudio}{0.11}

\newcommand{\julSixteenIndex}{\ensuremath{-}0.001}

\newcommand{\supFIndexDate}{2026-06-01}

\newcommand{\supFIndexJump}{0.34}

\newcommand{\tickJunAugAbs}{18}
\newcommand{\tickJunAugT}{\ensuremath{-}2.85}

\newcommand{\tickFullAbs}{10}
\newcommand{\tickFullT}{\ensuremath{-}2.69}

\newcommand{\twoTicketMinAbs}{22}

\newcommand{\twoTicketMaxAbs}{10}

\newcommand{\didIndex}{0.10}

\newcommand{\nCommits}{27}
\newcommand{\nSignoff}{14}
\newcommand{\timeObsSeven}{63}

\newcommand{\timeNullSeven}{67}

\newcommand{\timePSeven}{0.73}

\newcommand{\lagSimReps}{1,000}

\newcommand{\phaseN}{25}

\newcommand{\powerTwentyOne}{37}

\newcommand{\signoffLag}{7}

\newcommand{\trendAOnly}{0.124}

\newcommand{\trendAOnlyP}{$<$0.001}

\newcommand{\holmFive}{conversation structure}
\newcommand{\bhFive}{conversation structure, AV sync, audio quality, interview outcome}
\newcommand{\rawFive}{conversation structure, latency, AV sync, audio quality, interview outcome}

\newcommand{\nBhFive}{4}
\newcommand{\dedOutcome}{\ensuremath{-}0.8}

\newcommand{\dedOutcomeP}{0.02}
\newcommand{\dedLatencyBH}{0.06}
\newcommand{\twoIndexMin}{0.05}

\newcommand{\twoIndexMax}{0.18}

\newcommand{\twoCritMinAbs}{3.8}

\newcommand{\twoCritMaxAbs}{2.2}

\newcommand{\twoAnyMinAbs}{14.8}

\newcommand{\twoAnyMaxAbs}{0.9}
\newcommand{\costMonthly}{7,806}
\newcommand{\avoidedEndTech}{428}

\newcommand{\avoidedJuneBase}{1900}

\newcommand{\referenceMonthlyEnd}{1,800}
\newcommand{\reductionShareEnd}{23}

\newcommand{\costPerAvoidedTech}{18}

\newcommand{\creditHundredTech}{18}

\newcommand{\actualMonthlyEnd}{1,400}
\newcommand{\aprIndexPcl}{0.29}
\newcommand{\cadMin}{0}
\newcommand{\cadMax}{5}
\newcommand{\nProdDeploys}{3}

\newcommand{\rawChange}{0.13}

\newcommand{\thetaChange}{0.48}

\newcommand{\compChange}{0.34}

\newcommand{\thetaSd}{0.20}

\newcommand{\caseSd}{0.89}

\newcommand{\tickIntMean}{46}

\newcommand{\tickTechMean}{312}

\newcommand{\batchMin}{68}
\newcommand{\batchMax}{335}
\newcommand{\refreshIndexWeeksLo}{1}

\newcommand{\refreshIndexWeeksHi}{21}

\newcommand{\stgMayEleven}{8.50}

\newcommand{\stgMayEighteen}{6.83}

\newcommand{\stgJunOne}{5.77}

\newcommand{\stgJunEight}{4.10}

\newcommand{\stgJunTwentyNine}{2.92}

\newcommand{\stgJulThirteen}{6.08}

\newcommand{\stgAugSeventeen}{9.00}

\newcommand{\stgAugSeventeenN}{10}

\newcommand{\stgImprBefore}{8.5}

\newcommand{\stgImprTrough}{3.9}

\newcommand{\stgCohBefore}{1.00}

\newcommand{\stgCohTrough}{0.87}

\newcommand{\stgIdxBefore}{0.02}

\newcommand{\stgIdxTrough}{\ensuremath{-}0.11}

\newcommand{\stgOtherFamMinTrough}{0.92}

\newcommand{\stgNBefore}{116}
\newcommand{\stgNTrough}{78}

\newcommand{\swMar}{17}

\newcommand{\swMay}{27}

\newcommand{\refApr}{79}

\newcommand{\refAug}{44}

\newcommand{\volWeekMinK}{32}
\newcommand{\volWeekMaxK}{91}

\newcommand{\nJobs}{4,642}

\newcommand{\tickVolJunAugAbs}{14}
\newcommand{\tickVolJunAugT}{\ensuremath{-}2.68}

\newcommand{\tickVolFullAbs}{7}
\newcommand{\tickVolFullT}{\ensuremath{-}2.20}

\newcommand{\tickLevelFull}{3.7}

\newcommand{\tickLevelFullT}{1.43}

\newcommand{\phaseDropped}{2}

%% file: sections/01_intro.tex
\section{Introduction}\label{sec:intro}

A firm that puts an AI agent in front of its customers has to decide, day after day, whether the agent is working, whether to change it, and whether the evidence it already holds is still good. Evaluation programs exist to supply that evidence. An evaluation score is useful only if it still measures the same thing when the next operating decision arrives. For a deployed AI agent, that condition is easy to violate: the agent changes, the users change, and the human or model judges scoring the system may change too. A firm that compares raw scores across time can therefore mistake a new reviewer for a worse agent, or keep trusting an old evaluation after the deployed system has moved.

This creates a longitudinal measurement problem that differs from benchmark evaluation at a point in time. Production evaluation must satisfy three conditions. Scores must be comparable across batches despite changes in who rates them. They must remain current as the agent, its configuration, vendors, and user population drift. And they must track failures that matter outside the evaluation form. Existing work addresses important pieces separately: rater research shows systematic evaluator effects \citep{myford2009,casabianca2026,katoch2026,sachdeva2026}; longitudinal studies document model behavior changes across time \citep{chen2024drift,wiese2026,ford2026}; and production studies show that evaluation is central to operating AI systems \citep{pan2026map,gupta2026nubank}. What is less clear is how these problems interact when a firm uses repeated evaluations to infer whether one deployed agent is becoming more reliable.

We treat reliability as a latent state observed through noisy, rater-specific measurements. A simple decision model links three empirical objects to operating choices: rater severity determines whether scores can be compared over time; drift determines how long an evaluation remains informative; and the connection between evaluation findings, repairs, and market-facing outcomes determines whether the measured changes are operationally meaningful. The framework does not require a new estimator. Its purpose is to make explicit what a production evaluation program must identify before its score stream can support decisions.

We study a voice-and-video AI agent that conducts live job interviews. From March through August 2026, the firm's evaluation pipeline scored \nProd{} production recordings in \nBatches{} batches using a nine-family rubric. Reviewers overlapped within batches, which lets us separate reviewer severity from batch-level movement. We combine these ratings with a change log covering three production deploys and, from May, daily support-ticket data.

The evidence yields three findings. First, measurement composition can conceal improvement. Two reviewers scoring the same batches differed by \feRevBAbs{} standard deviations on the composite index. When the stricter reviewer joined, the raw trend flattened; after adjusting for reviewer effects, the latent index rose by \latentIndexChange{} standard deviations, while critical failures peaked at \latentCritPeak\% and then fell to \latentCritAug\%. Second, evaluated reliability moved materially even between logged deploys. At the estimated rate of unaccounted movement, uncertainty reaches the size of the largest observed deploy contrast after about five weeks, although this estimate is imprecise and driven by a small number of step-like changes. Third, an independent operational measure corroborates the direction of improvement: interview-related support tickets fell about \tickFullAbs\% per month relative to technical-issue tickets.

These results support three contributions. We make temporal comparability a first-order requirement for production evaluation; recast drift as the useful life of evaluation evidence; and separate whether an evaluation is informative from whether the evaluation program caused subsequent improvement. The last distinction is essential here: because the deployment has no unevaluated counterfactual, we do not identify the causal return to the evaluation program.

The rest of the paper is structured as follows. Section~\ref{sec:framework} develops the conceptual framework; Section~\ref{sec:setting} describes the agent, the evaluation program, and the data; Sections~\ref{sec:measurement} to~\ref{sec:decisions} present the evidence on comparability, drift, and operational relevance; Section~\ref{sec:conclusion} concludes.

%% file: sections/02_framework.tex
\section{Conceptual framework}\label{sec:framework}

This section writes down the problem of a firm that has deployed an AI agent and cannot observe directly whether it works. Nothing in the model is specific to interviewing; the agent could answer support queries, execute transactions or write code. The model is deliberately small. Its purpose is to derive, rather than assume, the objects the empirical sections estimate: a latent reliability state and the rate at which it moves, a measurement system whose readings contain the rater, a repair process with a cost and finite capacity, and a market-facing trace. Of the four propositions, two are estimated in this paper (the rater term in raw scores, Proposition~4, and the drift rate, Proposition~2) and two are not (Propositions~1 and~3, on how the value of evaluation runs through repair and repair capacity); the latter are the lens through which Section~\ref{sec:decisions} reads the evidence on the program's contribution. Derivations are in Appendix~\ref{app:model}.

\subsection{Environment}

Time is discrete and measured in weeks. In week $t$ the agent handles $V_t$ sessions. Its reliability is a state $\theta_t \in \mathbb{R}$; in general $\theta_t$ is a vector with one component per failure mode, and everything below applies component by component. A session that runs when the state is $\theta$ fails with probability $\phi(\theta)$, with $\phi' < 0$ and $\phi'' > 0$ over the operating range: reliability reduces failures at a diminishing rate. Each failure costs the firm $\lambda$, so the flow loss from unreliability is
\begin{equation}
\Lambda_t(\theta_t) = \lambda\, V_t\, \phi(\theta_t). \label{eq:loss}
\end{equation}

The state moves for two kinds of reasons. Engineers can change it deliberately: repair effort $e_t$, at cost $w$ per unit and bounded by capacity $\bar e$, raises the state by $g(e_t)$ with $g' > 0$, $g'' \le 0$; in practice, effort is a deploy of code, prompts or infrastructure, an intended repair whose realized effect is uncertain and can be negative. Everything else that moves the state is, from the firm's point of view, noise: configuration and environment churn that is not logged as a deploy, changes the vendor makes to a model served under a fixed name, and shifts in the population of sessions. \citet{lehman1980} observed that a system in use must change continually; we assume only that the state moves, and need a variance rather than a direction:
\begin{equation}
\theta_{t+1} = \theta_t + g(e_t) + \eta_t, \qquad \mathbb{E}[\eta_t] = 0,\quad \operatorname{Var}(\eta_t) = q, \label{eq:state}
\end{equation}
so that $q$ is the variance per week of the movement the firm's own records do not account for. Section~\ref{sec:drift} estimates $q$ with and without the deploys, which is the distinction between $g(e_t)$ and $\eta_t$.

The state is not observed. The firm can buy a reading of it by having raters review a sample of $n$ sessions from week $t$. Rater $r$ scores session $i$ as
\begin{equation}
y_{irt} = \theta_t + \gamma_r + \varepsilon_{irt}, \qquad \operatorname{Var}(\varepsilon_{irt}) = \sigma^2, \label{eq:meas}
\end{equation}
where $\gamma_r$ is the rater's severity, constant over time and specific to the failure mode, and $\varepsilon_{irt}$ collects the session's idiosyncratic difficulty and scoring noise. That ratings contain the rater is among the oldest results in measurement \citep{thorndike1920,cooper1981,linacre1989}, and severity differs across raters and drifts within them \citep{congdon2000,myford2009}. A batch costs $c(n) = c_0 + c_1 n$. If a share $s_{rt}$ of the batch is scored by rater $r$, the batch mean is
\begin{equation}
\bar y_t = \theta_t + \sum_r s_{rt}\,\gamma_r + \bar\varepsilon_t, \qquad \operatorname{Var}(\bar\varepsilon_t) = \sigma^2/n. \label{eq:batchmean}
\end{equation}

Finally, the firm's loss has a trace it does not choose to collect: users who experience a failure may complain. Complaints in week $t$ are $C_t = \pi_t\, V_t\, \phi(\theta_t) + u_t$, where $\pi_t$ is the propensity of a user who experienced a failure to file a report and $u_t$ is noise. Only some such users complain \citep{fornell1987}, so complaints are a selected signal of quality, and $\pi_t$ can move with the composition of users.

\subsection{The firm's problem}

The firm chooses an evaluation policy and a repair policy to minimize the expected discounted sum of the loss from unreliability, the cost of evaluation and the cost of repair:
\begin{equation}
\min_{\{h,\, n,\, e_t\}}\ \mathbb{E}\sum_{t} \beta^t \Big[ \Lambda_t(\theta_t) + w\,e_t + \mathbf{1}\{\text{batch in } t\}\, c(n) \Big], \qquad 0 \le e_t \le \bar e, \label{eq:problem}
\end{equation}
where $h$ is the interval between batches and the repair decision $e_t$ may depend on everything the firm has observed. The firm's information about $\theta_t$ is a posterior with mean $\hat\theta_t$ and variance $P_t$; between batches the variance grows as $P_t + q\,s$ after $s$ weeks, and a batch of $n$ sessions brings it to $(1/(P_t + q s) + n/\sigma^2)^{-1}$ \citep{harvey1989,durbin2012}. Repair is directed at the failure mode whose estimated state is furthest below its target.

Two objects are defined on this problem. The value of the information in a batch is the reduction it produces in the minimized expected loss, repair cost included, when the repair is chosen after the batch rather than before it; this is the general definition \citep{blackwell1953,raiffa1961} and needs no approximation. For the monitoring results we add a \emph{quadratic monitoring approximation}. Expanding the loss to second order around the posterior mean gives $\mathbb{E}[\Lambda_t(\theta)] \approx \Lambda_t(\hat\theta) + \kappa_t P$ with $\kappa_t = \lambda V_t \phi''(\hat\theta)/2$. If in addition the decision loss is quadratic in the gap between the repair chosen and the repair that would be optimal under full information, and repair is linear and unconstrained over the relevant range, the expected excess loss from acting on an estimate equals $\kappa_t P$, the certainty-equivalence property of linear-quadratic control, and the value of information reduces to $\kappa_t$ times the reduction in posterior variance. Propositions~1 and~2 invoke this approximation where stated; it is a monitoring rule, not a solution of \eqref{eq:problem}, whose failure loss is nonlinear and whose repair is costly and capacity-bound. We treat $\kappa_t$ as constant over the operating range.

\subsection{Propositions}

\paragraph{Proposition 1 (no value without a contingent repair).} If repair cannot respond to what evaluation reveals, because $\bar e = 0$ or because the repair schedule is fixed in advance, the expected loss from unreliability is the same with and without evaluation, whatever $n$ and $h$: the gross value of the program is zero and its net value is minus its cost. If repair does respond, the gross value of a batch is the reduction in minimized expected loss it produces, which under the quadratic monitoring approximation is $\kappa_t$ times the reduction in posterior variance. The evaluation budget and the repair budget therefore cannot be judged separately. (Proof in Appendix~\ref{app:model}.)

\paragraph{Proposition 2 (the optimal interval is a function of the drift rate).} Under the quadratic monitoring approximation, the average cost per week of evaluating every $h$ weeks with batches of $n$ is
\begin{equation}
\frac{c(n)}{h} + \kappa\Big(P_0(h,n) + \frac{q\,h}{2}\Big), \qquad P_0(h,n) = \frac{\sqrt{(qh)^2 + 4qh\,\sigma^2/n} - qh}{2}, \label{eq:interval}
\end{equation}
where $P_0$ is the steady-state posterior variance just after a batch (Appendix~\ref{app:model}). The optimal interval $h^*$ falls with the drift rate $q$ and the loss weight $\kappa$ and rises with the cost of a batch; when the sampling variance is small relative to the drift accumulated over an interval it has the closed form $h^* = \sqrt{2c(n)/(\kappa q)}$. This is the sampling-interval logic of the economic design of control charts \citep{duncan1956,lorenzen1986}, carried to a process that drifts rather than jumps; the costly, capacity-bound repair problem around it is our specialization. Its content for the present question is that ``evaluate continuously'' is the limit $q \to \infty$ or $c \to 0$; for a finite drift rate the recommendation has a number attached, and the number requires an estimate of $q$. The same calculation gives the horizon at which a reading becomes uninformative: the forecast variance $P_0 + qs$ reaches the size of a change the firm cares about, $D^2$, at $s = (D^2 - P_0)/q$. The horizon needs $q$, $P_0$ and $D$; the interval needs $c$ and $\kappa$ as well, which the data do not contain, so Section~\ref{sec:drift} reports the horizon and states the interval as the model's result. (Proof in Appendix~\ref{app:model}.)

\paragraph{Proposition 3 (the value of evaluation runs through what repair capacity lets the firm do with it).} The value of a finding is zero exactly when the optimal repair is the same under every finding. With no capacity that holds (Proposition~1). It also holds when capacity suffices for every mode and repair is costless, since everything is then repaired regardless; with costly repair the finding still tells the firm which repairs are worth making, so ample capacity leaves a selection value, smaller than under a binding constraint. In between, findings determine which modes are repaired and when: with capacity for one repair and two candidate modes, knowing which is worse is worth the difference in their shortfalls. The relationship is therefore not monotone in general, and evaluation and capacity are neither complements nor substitutes (Appendix~\ref{app:model} gives the two-mode example). The organizational learning literature makes the related point that failures must be surfaced and acted on for learning to occur \citep{sitkin1992,edmondson1999,madsen2010,tyre1997}. What is observable is the queue: findings that wait for capacity indicate that repair, and findings acted on as they arrive indicate that information, is the scarcer input at the margin. Section~\ref{sec:cadence} reads the changelog for that queue. (Proof in Appendix~\ref{app:model}.)

\paragraph{Proposition 4 (raw scores misestimate the agent when the roster changes).} Suppose the firm reads batch means \eqref{eq:batchmean} as readings of $\theta$. Between two batches $t$ and $t'$ its inferred change in the state is
\begin{equation}
\bar y_{t'} - \bar y_t = (\theta_{t'} - \theta_t) + \sum_r (s_{rt'} - s_{rt})\,\gamma_r + (\bar\varepsilon_{t'} - \bar\varepsilon_t), \label{eq:bias}
\end{equation}
so a change in who scores enters as a change in the agent, with sign and size set by the entering rater's severity times his share. Because $\gamma_r$ is specific to the failure mode, subtracting a single leniency constant leaves mode-level bias, and a firm that directs repair at the mode with the worst reading may repair the wrong one. The bias is removable under two conditions and one structural fact. The conditions are that sessions are assigned to raters within a batch independently of their difficulty, and that rater severity is constant over time. The structural fact is that the rater-by-batch graph is connected: within a batch the state is common to all raters, so severity differences are identified there, and a rater who scored two batches identifies the change in the state between them. This is the identification argument for worker and firm effects in connected sets of movers \citep{abowd1999,abowd2002} and, in spirit, for judge fixed effects under quasi-random assignment \citep{kling2006,dobbie2018}; Section~\ref{sec:measurement} states and partly tests the conditions for this setting, and shows that the roster term in \eqref{eq:bias} was about the size of the improvement itself. (Proof in Appendix~\ref{app:model}.)

\paragraph{The market trace.} Complaints per session are $C_t/V_t = \pi_t\,\phi(\theta_t) + u_t/V_t$, so a fall in complaints per session reflects a rise in reliability if the propensity to complain is stable, and the ratio of complaints about the agent to complaints in an unrelated category on the same instrument removes shocks common to the instrument. Neither identifies $\phi(\theta_t)$ if $\pi_t$ moves, which it may when the user population changes, so the model licenses a claim about direction, and Section~\ref{sec:outcomes} makes that claim and no stronger one.

%% file: sections/03_setting.tex
\section{Setting and data}\label{sec:setting}

\subsection{The agent and its deployment}

Zara is a voice-and-video AI agent that conducts live job interviews for micro1, a firm that recruits and vets professionals for AI-training and other remote work. A candidate joins a video call; the agent introduces itself, asks questions adapted to the role and to the candidate's answers, follows up, and closes the interview; the session is recorded. The agent runs on a real-time voice model for the conversation and a text model for reasoning and tool calls, orchestrated by a service the firm maintains. The firm's job-posting records imply roughly \volWeekMinK,000 to \volWeekMaxK,000 interviews a week over the study window, with volume close to doubling between March and September 2026 (Figure~\ref{fig:population}). Because the interviews are real, a failure in the agent is a failure in a candidate's job application.

We reconstructed the agent's change history from the two branches of its code repository, producing a changelog of dated commit groups classified by area, \nCommits{} of them inside the evaluation window of 26 March to 31 August. Two facts from it organize the analysis. First, the production branch received three behavior-relevant promotions of development work in the window, which we call the production deploys: 12 April (a new realtime model alias, a rewritten text-to-speech selector and a large prompt rewrite), 16 July (the next realtime model, region routing and prompt trims) and 29 July (June and July development work, including a reconnect timeout and a question tracker). One further production commit, on 20 August, moved credentials out of the repository and changed no behavior. A commit dates the code change rather than the moment it began serving traffic, and we take the two to coincide. Second, the production models ran under fixed version names, so the vendor's public launches during the window (23 April, 7 July and 9 July) changed nothing in production and serve as placebo dates. A fixed name is not a snapshot, since the vendor can update what it serves under the same name, and the changelog records the repository rather than configuration kept outside it; Section~\ref{sec:drift} returns to both limits.

\subsection{The evaluation pipeline}

From late March 2026 the firm had trainers watch recordings of completed production interviews and score them against a nine-family rubric: conversation structure, interruption handling, latency, speech pacing, coherence, audio-visual sync, freezes, audio quality and interview outcome. Each family contains several sub-items; we score a family as the share of its sub-items passed, on a 0 to 1 scale, and define the \emph{stability index} as the mean of the nine family scores, standardized. Trainers also give an overall impression on a 1 to 10 scale and flag a \emph{critical failure} when one of a short list of deal-breakers occurred (unrecoverable audio-visual desynchronization, data loss, session dropout, a persistent interruption pattern, and a few others); \avsyncShare\% of production critical failures are audio-visual sync failures. We also use a broader failure flag equal to one if the trainer either flagged a critical failure or gave an impression of 6 or below.

The pipeline scored \nProd{} production recordings between 26 March and 24 August 2026, in \nBatches{} batches keyed to the interview date, roughly fortnightly, of \batchMin{} to \batchMax{} recordings each; the last batch contains 35 because the data were pulled before it was complete, and we mark it as partial and exclude it from trend estimates. Each score carries the date the trainer signed it off, which for half of a batch is about \signoffLag{} days after the interview date; the interview date is the clock for the agent's state and the sign-off date the clock for when a finding existed. The pipeline also scored \nStaging{} recordings from pre-release channels on which the firm tested changes before deploying them. They are outside the main sample and appear once, in Section~\ref{sec:prerelease}, where the pre-release channel's own path is the object.

Three trainers scored production. \rev{A} scored every batch and scored alone until 21 May; \rev{B} joined on 21 May and scored roughly half of every batch thereafter; \rev{C} scored part of the 21 May batch. The roster, shown in the bottom panel of Figure~\ref{fig:latent}, is the identifying structure of the paper: because \rev{A} spans the whole window, the agent's path is identified on her scale, and because \rev{B} overlaps with her in nine batches, his severity is identified within batch. Two descriptive facts carry through. Most families are at ceiling most of the time: the median family scores full marks in more than 90\% of recordings, latency has the most headroom at \latencyAtMax\% at ceiling, and conversation structure is the family with the most frequent deductions, so movement in the index is movement in a few families. And raw critical-failure rates by batch range from zero to \obsCritPeak\%, with sampling standard errors of one to two percentage points, so single batches are noisy and the smoothing in Section~\ref{sec:measurement} does real work.

\subsection{Support tickets and job records}

A support-ticket dashboard launched on 4 May 2026 and records daily counts of tickets created by category. We use daily creation counts from 5 May to 17 September for interview-related tickets, which candidates file (mean \tickIntMean{} per day), and technical-issue tickets (mean \tickTechMean{} per day). Technical-issue tickets are the control: they are filed by the same users through the same instrument and share any shift in traffic, triage or category definitions that hits the dashboard as a whole, while having no particular reason to move with the interviewing agent's reliability.

\begin{figure}[t]
\centering
\includegraphics[width=\textwidth]{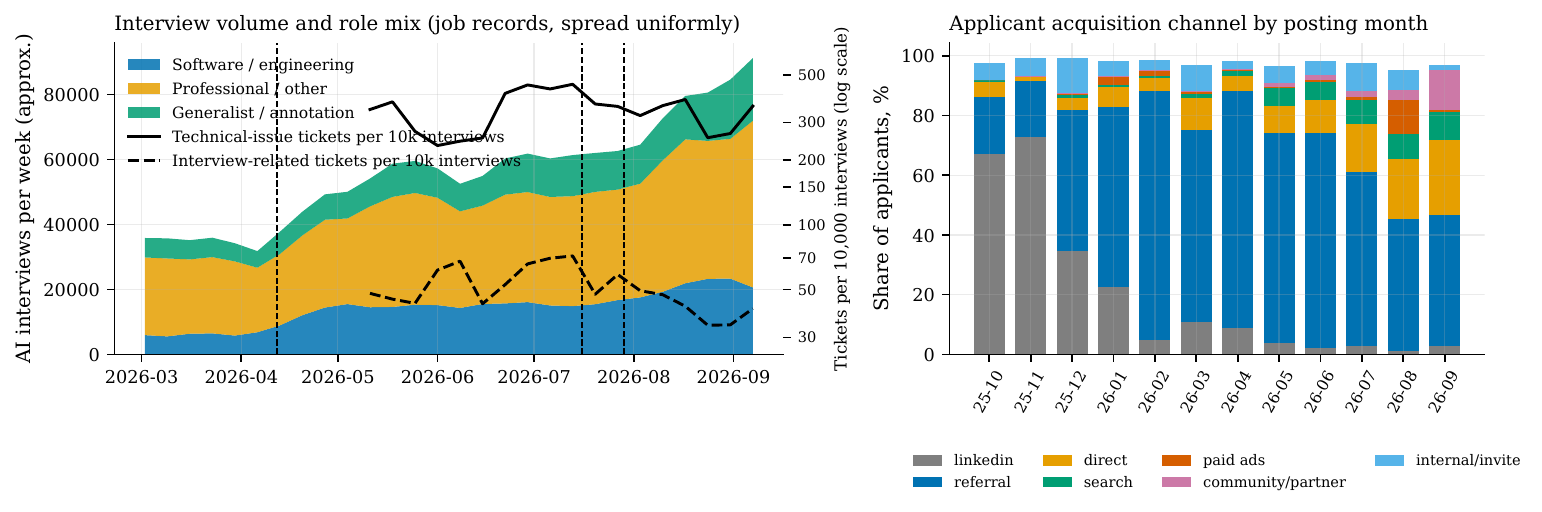}
\caption{The interview population during the window, from the firm's job-posting records. Left: approximate weekly AI-interview volume by role family (each posting's interviews spread uniformly over its active period), with support tickets per 10,000 interviews on the right axis from the dashboard's launch on 5 May, technical-issue (solid) and interview-related (dashed). The right axis is logarithmic: the tick labels are actual counts per 10,000 interviews, and equal vertical distances are equal percentage changes, so the two series can be compared despite their different levels. Dashed vertical lines mark production deploys. Right: applicants by acquisition channel, attributed to the posting month.}
\label{fig:population}
\end{figure}

The evaluated recordings carry no identifier for the job or the candidate, so nothing about the interview population is observed for the sample itself. The firm's job-posting records describe the population the sample was drawn from: \nJobs{} postings from September 2024 to 15 September 2026, each with its posting date, its applicants by acquisition source and the AI interviews it generated. Spreading each posting's interviews over its active period gives an imputed weekly volume and role mix, and attributing applicants to the posting month gives the acquisition channel over time (Figure~\ref{fig:population}). The volume series is an order-of-magnitude denominator, and the ticket analysis uses it to read tickets per interview. Two shifts matter downstream. The share of software and engineering roles among interviews rose from \swMar\% in March to \swMay\% in May and stayed near that level through August. The acquisition channel shifted from June: referral supplied \refApr\% of applicants to postings made in April and \refAug\% for August postings, while direct, search and paid traffic rose. Both bear on the assumption that the interview mix reaching the evaluation batches was comparable across batches, and Section~\ref{sec:measurement} weighs them where the adjusted path is interpreted.

\subsection{What is estimated, and how}\label{sec:empirical}

The deployment instantiates the primitives of Section~\ref{sec:framework}. The latent state is the batch effect net of reviewer effects, smoothed over batches; the measurement system is the rubric and the roster; the firm's own changes are the three deploys; unlogged movement is what remains between them; the repair process shows up as the cadence of commit groups; and the market trace is the ticket record. Table~\ref{tab:idmap} maps each quantity the paper estimates to its design, its assumptions and the result it serves, and marks the two results that are read as lenses rather than estimated.

\begin{table}[t]
\centering\small
\caption{Identification map: estimands, designs and what carries each inference}
\label{tab:idmap}
\begin{tabular}{p{3.4cm}p{4.2cm}p{5.2cm}p{2.2cm}}
\toprule
Estimand & Design & Key assumptions & What it delivers \\
\midrule
Rater severity $\gamma_r - \gamma_{r'}$ (Proposition~4) & Within-batch comparison of raters; rater FE with batch FE & Assignment within batch unrelated to interview difficulty; additivity & Partly (per-batch gap stability) \\
Latent path $\theta_t - \theta_1$ & Batch FE net of rater FE (Stage 1); local level with estimated $q$ (Stage 2) & Comparable interview mix across batches; rubric stable within the window; additive rater effects & Partly (role mix and acquisition channel from job records; additivity) \\
Where the improvement sits & Family-level paths and trends & As above, per family; ceiling effects noted & Descriptive \\
Deploy vs vendor coincidence & Two-batches-before / two-after contrasts with rater FE & No other change at the same date; fixed version names (documented); silent vendor updates not excluded & Placebo dates available \\
Drift rate $q$ (Proposition~2) & Profile ML on Stage-1 batch effects, with and without free steps at deploys & Local-level dynamics; correct Stage-1 sampling variances & LR test of $q=0$ \\
Market agreement & Ticket ratio trend against same-instrument control and against interview volume; two-period splits & Control category shares instrument shocks and not interview-specific ones; stable propensity to complain & Alternative controls reported \\
Program effect on reliability and outcomes (Proposition~1, lens) & Deploy contrasts, pre-release episode, ticket agreement & A causal magnitude needs a no-evaluation comparison or exogenous variation in evaluation intensity; neither exists here & Direction only; magnitude not identified \\
Repair capacity (Proposition~3, lens) & Commit-group cadence against the cadence of findings & Repository activity records work, not capacity or backlog & Descriptive \\
\bottomrule
\end{tabular}
\end{table}

%% file: sections/04_improving.tex
\section{Can the firm tell whether the agent is improving?}\label{sec:measurement}

Changing measurement can conceal progress. This section shows how much of a score is the rater, estimates the agent's path with the rater removed, and asks where in the rubric the movement sits and when it arrived. It opens with the specifications it estimates.

\subsection{Specification}\label{sec:method}

Equation~\eqref{eq:meas} of Section~\ref{sec:framework} is the measurement equation: a score is the agent's state in the batch, plus the reviewer's severity, plus noise. We estimate the agent's latent path from it in two stages, and we describe each stage and what it assumes because the description is part of the contribution: a firm that adopts the method should know what it is buying.

\paragraph{Stage 1: reviewer-adjusted batch effects.} For each outcome (the index, each family's score or deduction indicator, the critical-failure flag, the impression) we estimate equation~\eqref{eq:meas} on production recordings by least squares with a full set of batch dummies and reviewer dummies, omitting the dummy for \rev{A}. The batch coefficients $\hat\theta_t$ are the outcome's expected value in batch $t$ for an interview scored by \rev{A}; nothing is assumed about how $\theta_t$ moves over time. What is assumed is (A1) that the mix of interviews reaching each batch is comparable across batches, (A2) that the rubric and the trainers' interpretation of it are stable within the window, and (A3) that reviewer effects are additive, which Section~\ref{sec:rater} tests and the reviewer-A-only trend sidesteps. A1 cannot be tested on the recordings, which carry no covariates about the job, the candidate's device or the network; the job records offer two partial checks, taken up where the adjusted path is interpreted. A2 is maintained, plausible for a rubric that did not change and trainers who did not turn over. The sampling covariance of the $\hat\theta_t$ is heteroskedasticity-robust, with a floor on the variance of the rare critical-failure outcome so that a batch with zero events is not treated as infinitely precise, and it is not diagonal, because batches scored by both reviewers share the estimated offset $\hat\gamma_B$; Stage 2 carries the full covariance, which moves the smoothed first-to-last change in the index from \latentIndexChangeDiag{} to \latentIndexChange{} standard deviations (Appendix~\ref{app:est}).

\paragraph{Stage 2: local-level smoothing.} The Stage-1 estimates are noisy readings of a state that moves. We fit
\begin{equation}
\hat\theta_t = \mu_t + v_t,\qquad \mu_{t+1} = \mu_t + \eta_t,\qquad (v_1,\dots,v_T) \sim N(0,\hat\Sigma),\quad \eta_t \sim N(0, q\,\Delta_t),
\end{equation}
where $\hat\Sigma$ is the Stage-1 sampling covariance of the batch effects (known; its off-diagonal terms come from the shared reviewer coefficients) and $q$ is the innovation variance per week, estimated by profile maximum likelihood with a diffuse initial state. The smoothed $\hat\mu_t$ is the latent path; the smoother weights each batch by its precision and borrows strength from neighbors in proportion to how slowly the state is estimated to move, and with a diagonal $\hat\Sigma$ it is the Kalman smoother. If $q$ were zero the path would be a constant; if $q$ were very large the path would reproduce the Stage-1 estimates. The estimated $q$ is the drift rate of Section~\ref{sec:drift}, and a likelihood-ratio test of $q=0$ asks whether the state moved at all beyond sampling noise.

\paragraph{Trends and contrasts.} Trends are estimated by regressing the case-level outcome on time in months with reviewer fixed effects, with standard errors clustered on batch day and a wild cluster bootstrap check; one specification adds deploy-regime fixed effects so that the trend is identified from within-regime movement only, and one uses \rev{A}'s cases alone. Family trends use each family's deduction indicator and report Holm and Benjamini-Hochberg adjustments across the nine families tested. Deploy contrasts compare the two production batches after a date with the two before, with reviewer fixed effects, and are read as magnitudes rather than tests.

\paragraph{What the estimator delivers and does not.} The output is a path of the agent's reliability on \rev{A}'s scale, with the reviewer composition removed and sampling noise reduced. Levels are relative: we cannot say whether \rev{A} or \rev{B} is ``right'' about the agent, only that the movement over time is the same on either scale. The path captures the agent through the rubric, so a family the rubric does not cover is invisible and a recording failure that reviewers score as an agent failure enters as one, and it captures whatever moves with time and is common to all reviewers, which is the agent plus the interview population (A1).

\subsection{How much of a score is the rater}\label{sec:rater}

Table~\ref{tab:rater_fe} reports rater fixed effects and a variance decomposition. Batch fixed effects, which carry everything the agent did from one fortnight to the next, explain \rTwoBatch\% of case-level variance in the production index; adding trainer fixed effects raises this to \rTwoBatchTrainer\%. Which reviewer scored a recording predicts its score far better than which fortnight it came from, because the agent's between-batch movement (a standard deviation of \thetaSd{} across batches) is small next to both the reviewer offset and the interview-to-interview noise (case-level standard deviation \caseSd). The trainer who joined in May, \rev{B}, scored the same batches \feRevBAbs{} standard deviations (SE \feRevBSe) below \rev{A} on the index, and flagged critical failures at \critRevB\% against \rev{A}'s \critRevA\%, a within-batch difference of \feRevBCrit{} percentage points in the other direction. A reviewer can be stricter on graded items and more lenient on the deal-breaker flag at the same time.

\input{tables/tab_reviewer_effects}

\begin{figure}[t]
\centering
\includegraphics[width=\textwidth]{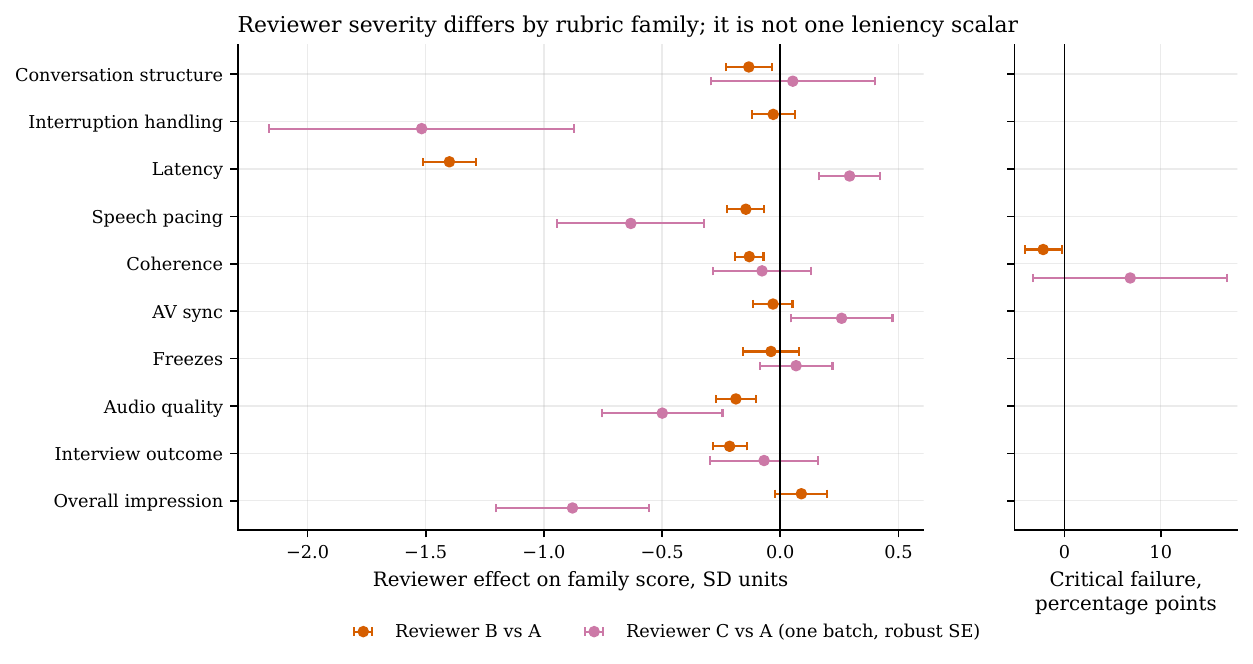}
\caption{Reviewer effects by rubric family, production recordings, batch fixed effects, relative to trainer A, with 95\% confidence intervals (standard errors clustered on batch day; heteroskedasticity-robust for trainer C, who scored production in one batch of 47 recordings).}
\label{fig:severity}
\end{figure}

Figure~\ref{fig:severity} shows why a single leniency adjustment would not work. Estimating the \rev{B} effect family by family, with batch fixed effects, gives \sevLatencyRevB{} standard deviations on latency and \sevStructureRevB{} on conversation structure; the two trainers agree on audio-visual sync, freezes and interruption handling and disagree on latency, pacing and audio quality. Severity is a vector with one entry per family: it attaches to families whose sub-items ask for a judgment of degree (how long a pause is too long) and is close to zero for families whose sub-items record an event (did the video freeze), as in the rater-effects literature \citep{myford2009,leckie2011}. The reviewer-adjusted path must therefore be estimated family by family.

Equation~\eqref{eq:meas} assumes that a reviewer's severity is the same in every batch. Table~\ref{tab:gaps} tests this in the only way the data allow, by computing the \rev{B} minus \rev{A} gap on the index separately in each of the \gapN{} batches both scored. The gap is negative in every batch and ranges from \gapMin{} to \gapMax{} standard deviations; the inverse-variance pooled gap is \gapPooled{} (SE \gapPooledSe), and a heterogeneity test rejects exact constancy ($Q=\gapQ$, $p=\gapQp$). Additivity is therefore an approximation, a good one for the direction and rough size of movement over months, since the variation in the gap is small relative to the gap itself and shows no trend, and a poor one for any question that turned on a single batch's level, which we do not ask. The direct check drops the adjustment altogether: reviewer A scored every batch, so a trend on her scores alone needs no reviewer effect, and Table~\ref{tab:trends} gives \trendAOnly{} standard deviations per month ($p=\trendAOnlyP$) against \trendFe{} with the adjustment. A common drift in both reviewers' standards would survive this check and is indistinguishable from movement in the agent; we hold it as the maintained assumption A2.

\input{tables/tab_reviewer_gaps}

The practical rule follows directly. Overlap raters, and never read a raw average across a changing roster: a program whose raters do not overlap cannot separate its raters from its agent, and one that reads raw averages will misread its agent at exactly the moments the roster changes. The fix is a matter of assignment: a fraction of every batch to two raters, at least one rater on every batch, and reviewer effects estimated within batch before any trend is reported. The same applies to model judges, whose severity is documented \citep{katoch2026,sachdeva2026} and whose behavior changes across versions \citep{chen2024drift}; a judge version is a rater.

\subsection{Raw path versus adjusted path}\label{sec:paths}

\begin{figure}[t]
\centering
\includegraphics[width=0.92\textwidth]{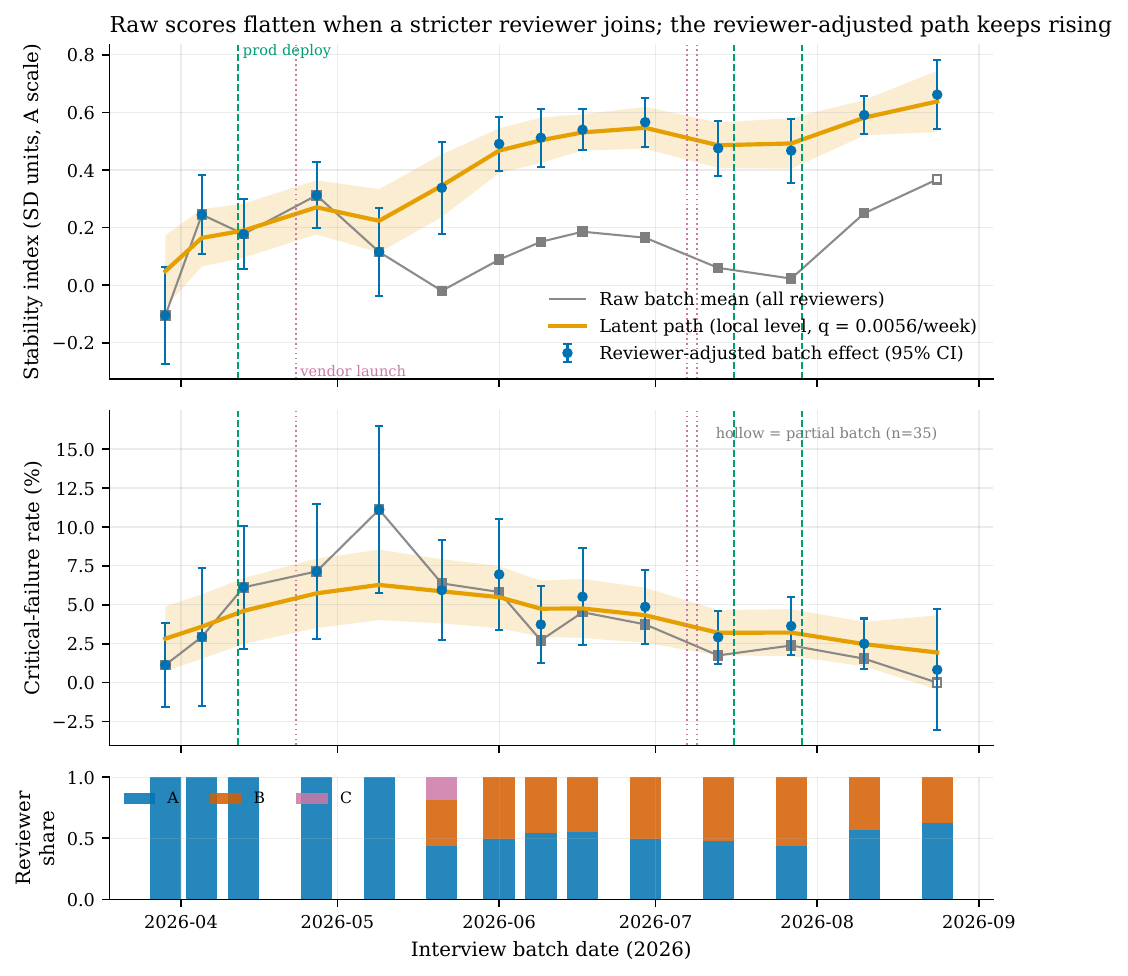}
\caption{Raw batch means, reviewer-adjusted batch effects and the smoothed latent path for the production stability index (top) and the critical-failure rate (middle), with the reviewer roster by batch (bottom). Dashed green lines mark the three production deploys; dotted purple lines mark vendor model launches, on which nothing in production changed. The 24 August batch is partial (35 recordings).}
\label{fig:latent}
\end{figure}

Proposition~4 of Section~\ref{sec:framework} says that a raw batch mean equals the state plus a roster term. Figure~\ref{fig:latent} shows the result for the index and the critical-failure rate: the gray line is the raw batch mean across all reviewers, the blue points are the Stage-1 reviewer-adjusted batch effects with 95\% intervals, and the orange line is the smoothed latent path. The two diverge exactly where the roster changes, and the divergence is arithmetic: each raw batch mean equals the adjusted batch effect plus a composition term, the share of the batch scored by each reviewer times that reviewer's offset. From 9 May to 10 August the raw mean rose \rawChange{} standard deviations while the adjusted effect rose \thetaChange{} and the composition term fell \compChange{}: the agent's improvement and the roster change are of the same size and arrive together, which is why they cancel in the raw series. Table~\ref{tab:trends} puts numbers on it: the raw trend in the production index is \trendRaw{} standard deviations per month ($p=\trendRawP$), the reviewer-adjusted trend is \trendFe{} (SE \trendFeSe, $p\trendFeP$, wild cluster bootstrap $p=\trendFeBoot$), and adding deploy-regime fixed effects gives \trendRegime{} ($p=\trendRegimeP$).

\input{tables/tab_trends}

The smoothed paths give the headline numbers. The latent index rises from \latentIndexFirst{} standard deviations on 29 March to \latentIndexAug{} on 10 August, the last complete batch, a change of \latentIndexChange{} (SE \latentIndexChangeSe). The latent critical-failure rate does something the index does not: it starts at \latentCritFirst\%, rises to a peak of \latentCritPeak\% around \latentCritPeakDate, and falls steadily to \latentCritAug\% on 10 August. The raw rate in the partial August batch is \obsCritLast\% on 35 recordings, the figure the firm's own summary showed, which the smoothed path treats as one noisy reading among fourteen. The critical-failure series is a hump, and a one-number summary such as a trend misses the shape. The impression score, the trainer's overall 1 to 10 judgment, drifts down over the window on \rev{A}'s scale even as the index rises; it is the one item on the form that invites a holistic judgment, and the family-level evidence, where engineering work should show, is where we look for the agent.

One qualification belongs here. The adjusted path captures the agent plus the interview population (A1), and the recordings carry nothing about the population. The job records show that it moved on two dimensions: the share of software roles among interviews rose from \swMar\% to \swMay\% between March and May and then held, and the acquisition channel shifted from June, away from referral and toward direct, search and paid traffic. The first shift overlaps the spring rise in critical failures and is a candidate explanation for part of it; the second overlaps the summer improvement in the index, and whether a candidate's acquisition channel affects the technical behavior of a call is something these data cannot say. Two features of the evidence make a population explanation of the whole path less likely without excluding it: the movement is concentrated in the families that map to specific engineering work in the changelog (Section~\ref{sec:dynamics}), and it agrees in direction with a market signal that does not pass through the reviewers or the recordings (Section~\ref{sec:outcomes}).

\subsection{Which families moved}\label{sec:dynamics}

\begin{figure}[!htbp]
\centering
\includegraphics[width=\textwidth]{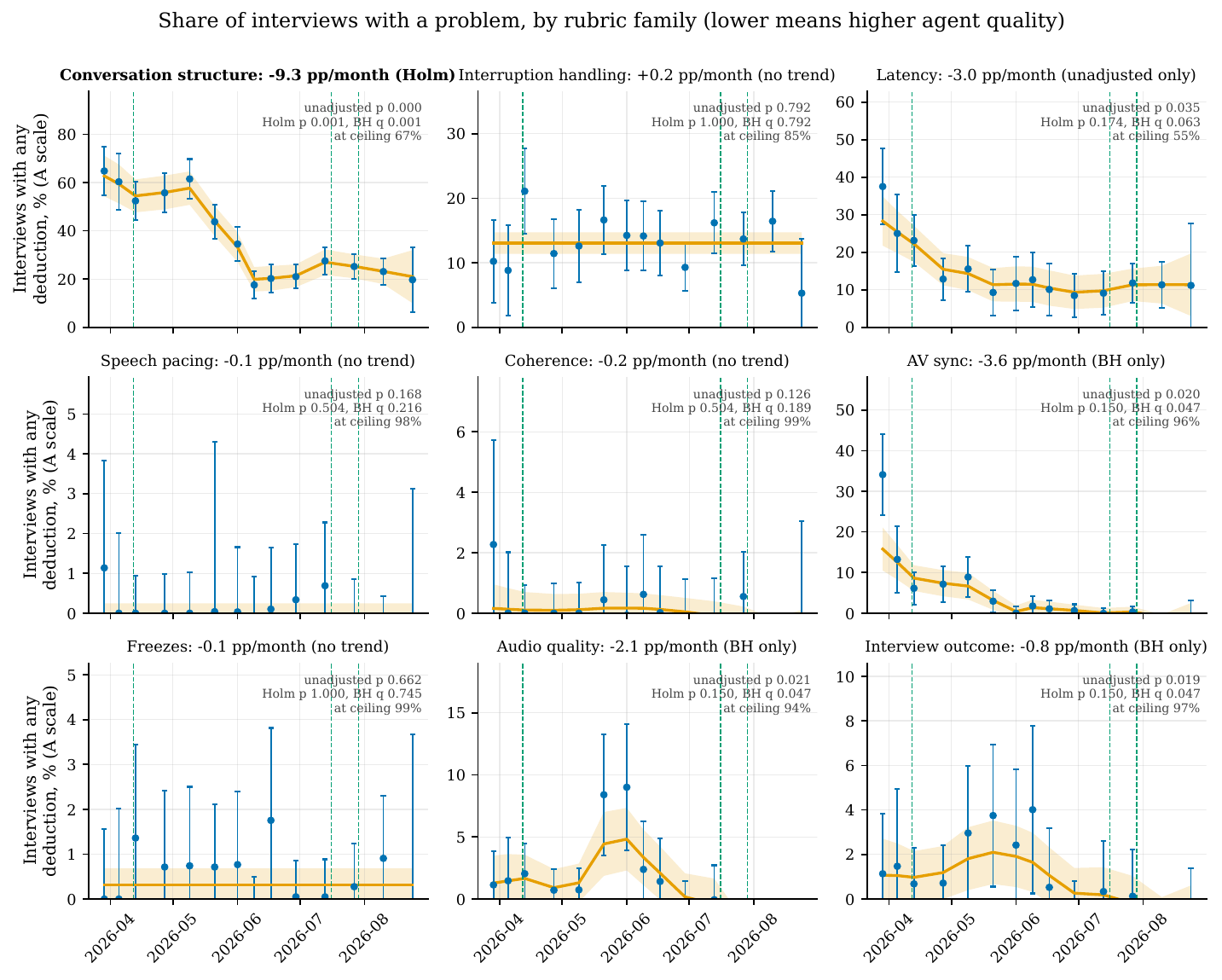}
\caption{Share of production interviews with a problem in each rubric family (a lost sub-item on the family's checklist), reviewer-adjusted. Lower means higher agent quality, so a negative trend is an improvement. Points are Stage-1 batch effects with 95\% intervals, the line and band the smoothed latent path; green dashed lines mark production deploys. Each panel title gives the family's trend in the deduction rate, in percentage points per month, from a regression of the deduction indicator on time with reviewer fixed effects (standard errors clustered on batch day, partial 24 August batch excluded), and the verdict of the nine-way multiplicity correction: ``Holm'' means the trend survives the Holm step-down at 5\% (conversation structure only, in bold); ``BH only'' means it survives the Benjamini-Hochberg adjustment at 5\% but not Holm (audio-visual sync, audio quality, interview outcome); ``unadjusted only'' means $p<0.05$ before correction and above it after (latency); ``no trend'' means $p\ge0.05$ (interruption handling, speech pacing, coherence, freezes). The corner of each panel reports the unadjusted $p$, the Holm $p$, the BH $q$ and the share of recordings at ceiling, which bounds how far a family's deduction rate can fall. Appendix Table~\ref{tab:famtrends} reports the standardized-score trends as well.}
\label{fig:families}
\end{figure}

Figure~\ref{fig:families} shows the reviewer-adjusted deduction rate, the share of interviews with at least one lost sub-item, for each of the nine families, with each family's monthly trend and its p-value adjusted for the nine families tested in the panel title (Appendix Table~\ref{tab:famtrends} gives the full regressions). The profiles differ in size more than in sign. Conversation structure went from \structDedFirst\% of interviews with a deduction to \structDedAug\% (a trend of \dedStructure{} percentage points per month, $p\dedStructureP$), with most of the drop between 9 May and 1 June. Audio-visual sync fell from \avsyncDedFirst\% to close to zero (\dedAvsync{} points per month, $p=\dedAvsyncP$), the decline beginning in April. Latency fell from \latencyDedFirst\% to \latencyDedAug\% (\dedLatency{} per month, $p=\dedLatencyP$), almost entirely before June. Audio quality followed a different shape: a spike to \audioDedPeak\% of interviews in late May and early June, then a fall to zero by July (\dedAudio{} per month, $p=\dedAudioP$). Interview outcome, deducted in a few percent of interviews, also declined (\dedOutcome{} points per month, $p=\dedOutcomeP$). Interruption handling, speech pacing, coherence and freezes show no movement. Five families have unadjusted $p<0.05$ (\rawFive); with the Benjamini-Hochberg adjustment across nine families \nBhFive{} remain (\bhFive; latency's adjusted $q$ is \dedLatencyBH), and with the Holm adjustment only \holmFive{} does. The conservative statement is that conversation structure fell by a large and unambiguous amount, that audio-visual sync, latency, audio and outcome fell by smaller amounts that a nine-way correction leaves uncertain individually, and that four families did not move. We use the composite index and the critical-failure rate, which are fixed in advance, wherever a summary is needed downstream, and refer to individual families only descriptively.

The families whose deduction rates fell are the families the changelog says were worked on. The 12 April deploy rewrote the text-to-speech selector and deleted several hundred lines of prompt; conversation structure and audio-visual sync improved through April and May. The June development-branch commits are about reconnection and stuck-session repair, the 29 July deploy shipped the reconnect timeout and the question tracker, and structure and audio step again at the end of July. The families that did not move are the ones no commit addresses. The correspondence is descriptive, and it is the kind a shift in the candidate population would be unlikely to produce on its own.

Two qualifications attach. Audio-visual sync accounts for \avsyncShare\% of critical failures, and the recording system had a defect that bears on it: a development-branch commit on 9 August notes that a candidate reconnecting mid-interview had started a second recording that overwrote the first. Reconnect handling dates from March, so the defect may have operated through much of the window, but the changelog does not date its onset; a reviewer watching a recording overwritten mid-session would plausibly score audio-visual sync or data loss as failed, so some critical failures may be failures of the recording through this path, in a share the records do not fix. (A second recording defect, in the screen-share stream, belongs to a structure introduced on the development branch in August that had not reached production by the end of the window.) And several families are at ceiling in more than 95\% of recordings, so their flatness is partly a floor on the deduction rate rather than evidence that nothing happened.

The rule for reporting follows. Report families, not indices. A composite index over nine weakly correlated families does not show which families moved and which did not, and does not show that a deploy improved structure while worsening deal-breakers; engineering acts on families, and the index serves a different reader.

\subsection{Changes around deploys and around the vendor launch}\label{sec:deploys}

\begin{figure}[t]
\centering
\includegraphics[width=\textwidth]{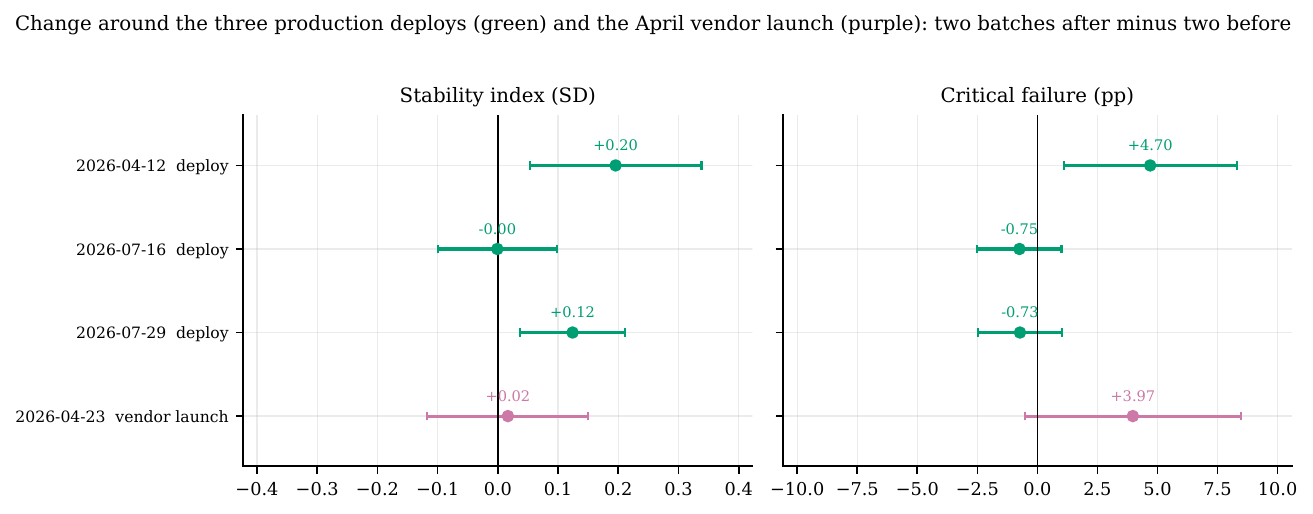}
\caption{Change in the reviewer-adjusted index and critical-failure rate around the three internal deploys (green) and the 23 April vendor launch (purple): the mean of the two production batches after the date minus the mean of the two before, with reviewer fixed effects. Intervals are 95\%, heteroskedasticity-robust, and too narrow by construction (Table~\ref{tab:deploys}); the contrasts are read as magnitudes against each other.}
\label{fig:deploys}
\end{figure}

Table~\ref{tab:deploys} and Figure~\ref{fig:deploys} compare the two production batches after each event with the two before, with reviewer fixed effects. Four batches per contrast leave no good basis for inference (for the April index contrast, $p=\aprIndexP$ with robust standard errors against $p=\aprIndexPcl$ clustered on the four batches), so we read the contrasts as magnitudes against each other, and as coincidence in timing rather than identified deploy effects. At the 12 April deploy the index rose by \aprIndex{} standard deviations and conversation structure by \aprStruct{}, while the critical-failure rate rose by \aprCrit{} percentage points and the impression fell by \aprImpr: a deploy that improved structured behavior and worsened deal-breakers at the same time, which is what the hump in Figure~\ref{fig:latent} shows. One reading is that a large deploy (a new realtime alias and a rewritten audio component) introduced a new failure mode while fixing old ones; the recording defect described above is the other candidate, and the scores cannot separate them. At the 29 July deploy the index rose by \julIndex, with structure ($+\julStruct$) and audio ($+\julAudio$) carrying it, and the critical-failure rate did not move detectably; the 16 July deploy shows nothing on the index (\julSixteenIndex).

\input{tables/tab_deploy_contrasts}

The placebo tests one thing: whether the scores moved on the date a new vendor version became publicly available, which under fixed version names should be inert (a vendor update served behind the same name at some other time would appear as unaccounted movement in Section~\ref{sec:drift}). Only the 23 April launch can serve, since the 7 and 9 July launches share every batch of their windows with the 16 July deploy. It shows an index change of \vendIndex{} and an impression change of \vendImpr, against \aprIndex{} and \aprImpr{} at the deploy eleven days earlier; its critical-failure contrast, \vendCrit{} points, is not clean, because the 13 and 27 April batches sit in both windows. Where the two windows separate, the vendor date shows nothing and the deploy date shows the movement.

\begin{figure}[t]
\centering
\includegraphics[width=0.75\textwidth]{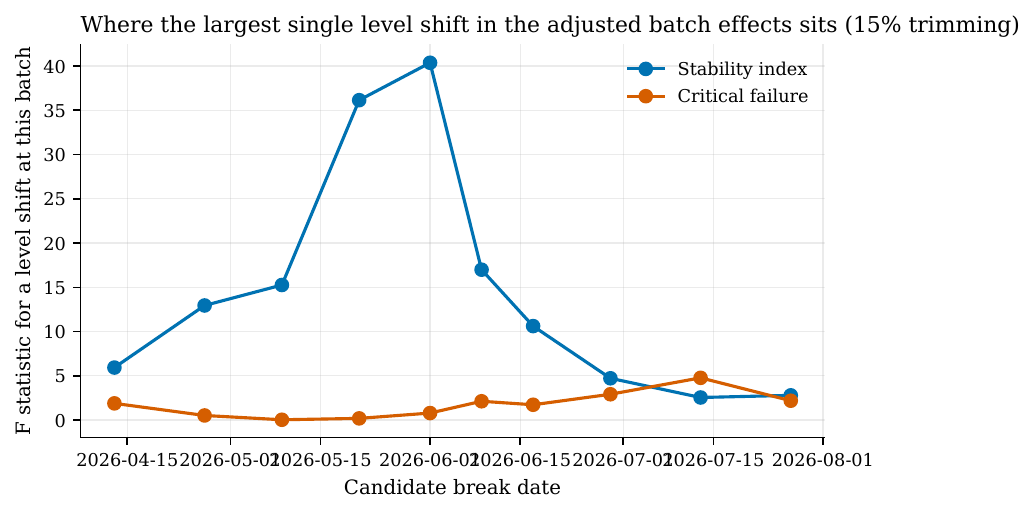}
\caption{Sup-$F$ statistics for a single level shift in the reviewer-adjusted batch effects at each candidate date (15\% trimming), used to locate the largest shift; with fourteen estimated points the asymptotic critical values are not applicable and none is drawn.}
\label{fig:supF}
\end{figure}

Locating the single largest level shift in the reviewer-adjusted index (Figure~\ref{fig:supF}, the sup-$F$ statistic of \citet{andrews1993} used descriptively, since its asymptotic critical values do not apply to fourteen estimated points) places it at \supFIndexDate{}, a jump of \supFIndexJump{} standard deviations, between the 21 May and 1 June batches, seven weeks after the last production deploy and in the middle of a run of development-branch work on reconnection. Either the April deploy took several batches to show fully, or a change reached production without a commit on the production branch, or the interview population shifted; the changelog cannot distinguish these. The critical-failure rate has no comparable single shift; its path is a gradual hump.

%% file: tables/tab_reviewer_effects.tex
\begin{table}[t]
\centering
\small
\caption{How much of a score is the rater: reviewer fixed effects and variance decomposition}
\label{tab:rater_fe}
\resizebox{\ifdim\width>\linewidth\linewidth\else\width\fi}{!}{%
\begin{tabular}{lrrr}
\toprule
 & Stability index (SD) & Critical failure (share) & Impression (SD) \\
\midrule
Reviewer B vs A & -0.792 (0.045)*** & -0.022 (0.010)** & +0.089 (0.056) \\
Reviewer C vs A (one batch) & -0.310 (0.128) & +0.068 (0.052) & -0.879 (0.166) \\
\midrule $R^2$, batch FE only & 0.014 & 0.016 & 0.051 \\
$R^2$, batch + reviewer FE & 0.162 & 0.021 & 0.081 \\
Cases & 2,611 & 2,611 & 2,611 \\
\bottomrule
\end{tabular}
}
\begin{minipage}{0.95\linewidth}\vspace{2pt}\footnotesize\emph{Notes:} Production cases, batch fixed effects, standard errors clustered on batch day (14 clusters); reference rater is reviewer A. Reviewer C scored production in a single batch of 47 recordings; his coefficients are identified from that batch alone, carry heteroskedasticity-robust standard errors, and are reported without significance markers. $^{*}p<0.10$, $^{**}p<0.05$, $^{***}p<0.01$.\end{minipage}
\end{table}

%% file: tables/tab_reviewer_gaps.tex
\begin{table}[t]
\centering
\small
\caption{Within-batch reviewer gap, B minus A, by production batch}
\label{tab:gaps}
\resizebox{\ifdim\width>\linewidth\linewidth\else\width\fi}{!}{%
\begin{tabular}{lrrrrrr}
\toprule
Batch & Index gap (SD) & SE & Critical-failure gap (pp) & SE & n B & n A \\
\midrule
2026-05-21 & -0.66 & (0.18) & -7.2 & (2.8) & 95 & 109 \\
2026-06-01 & -0.53 & (0.09) & -5.5 & (3.4) & 96 & 93 \\
2026-06-09 & -0.76 & (0.11) & -0.6 & (2.4) & 84 & 100 \\
2026-06-17 & -0.89 & (0.07) & -6.1 & (2.7) & 89 & 110 \\
2026-06-29 & -0.90 & (0.09) & -2.1 & (2.2) & 149 & 145 \\
2026-07-13 & -0.89 & (0.10) & +0.5 & (1.5) & 150 & 136 \\
2026-07-27 & -0.89 & (0.11) & -0.6 & (1.7) & 188 & 147 \\
2026-08-10 & -0.68 & (0.06) & +0.4 & (1.6) & 112 & 148 \\
2026-08-24 (partial) & -0.74 & (0.12) & +0.0 & (no events) & 13 & 22 \\
\midrule Pooled (inverse-variance) & -0.77 & (0.03) &  &  &  &  \\
Heterogeneity Q (df, p) & 19.0 & (8, 0.015) &  &  &  &  \\
\bottomrule
\end{tabular}
}
\begin{minipage}{0.95\linewidth}\vspace{2pt}\footnotesize\emph{Notes:} Each row compares the two reviewers on interviews from the same batch date, so the gap cannot reflect agent change over time: the index gap is reviewer B's mean score minus reviewer A's in standard-deviation units, and the critical-failure gap is the difference in their critical-failure rates in percentage points, each with its standard error. A stable gap supports an additive reviewer effect; the heterogeneity test asks whether the index gap varies across batches by more than sampling noise would produce.\end{minipage}
\end{table}

%% file: tables/tab_trends.tex
\begin{table}[t]
\centering
\small
\caption{Monthly trend in production reliability, March to August 2026}
\label{tab:trends}
\resizebox{\ifdim\width>\linewidth\linewidth\else\width\fi}{!}{%
\begin{tabular}{lrrrrrr}
\toprule
Outcome & Raw trend / month & Reviewer FE & Wild-boot p & Reviewer A only & Reviewer + deploy-regime FE & Reviewer FE, incl. partial batch \\
\midrule
Stability index (SD) & +0.002 (0.027) & +0.102 (0.021)*** & 0.002 & +0.124 (0.023)*** & +0.125 (0.032)*** & +0.102 (0.020)*** \\
Critical failure (pp) & -1.160 (0.450)** & -0.784 (0.451) & 0.193 & -0.967 (0.571) & -1.498 (0.606)** & -0.834 (0.431)* \\
Any failure flag (pp) & -1.957 (0.742)** & -0.965 (0.638) & 0.193 & -1.123 (0.806) & -2.699 (0.539)*** & -1.064 (0.626) \\
Impression (SD) & +0.059 (0.034) & +0.034 (0.033) & 0.371 & +0.037 (0.041) & +0.135 (0.035)*** & +0.033 (0.032) \\
\bottomrule
\end{tabular}
}
\begin{minipage}{0.95\linewidth}\vspace{2pt}\footnotesize\emph{Notes:} Production cases (2,576; last column adds the partial 24 August batch). Coefficients on time in months; standard errors clustered on batch day (13 clusters). 'Wild-boot p' is a restricted wild cluster bootstrap p-value (Webb weights, 999 draws) for the reviewer-FE trend. 'Reviewer A only' uses the scores of the one reviewer present in every batch and needs no reviewer adjustment; it is the check on additivity. Deploy regimes split the window at the three production deploys (12 April, 16 July, 29 July), so the last column identifies the trend from within-regime movement only. $^{*}p<0.10$, $^{**}p<0.05$, $^{***}p<0.01$.\end{minipage}
\end{table}

%% file: tables/tab_deploy_contrasts.tex
\begin{table}[t]
\centering
\footnotesize
\caption{Change in reviewer-adjusted scores across internal production deploys and vendor model launches}
\label{tab:deploys}
\resizebox{\ifdim\width>\linewidth\linewidth\else\width\fi}{!}{%
\begin{tabular}{lrrrrrr}
\toprule
Event & Index (SD) & Crit. fail (pp) & Impression (SD) & Structure (SD) & AV sync (SD) & Audio (SD) \\
\midrule
2026-04-12 Internal deploy & +0.20 (0.07) & +4.70 (1.84) & -0.43 (0.07) & +0.37 (0.09) & +0.31 (0.17) & +0.00 (0.02) \\
2026-07-16 Internal deploy & -0.00 (0.05) & -0.75 (0.89) & -0.09 (0.04) & +0.11 (0.06) & +0.03 (0.03) & +0.05 (0.04) \\
2026-07-29 Internal deploy & +0.12 (0.04) & -0.73 (0.90) & -0.05 (0.04) & +0.32 (0.05) & +0.03 (0.01) & +0.11 (0.03) \\
2026-04-23 Vendor launch (placebo) & +0.02 (0.07) & +3.97 (2.30) & -0.05 (0.07) & +0.05 (0.08) & -0.22 (0.17) & -0.01 (0.01) \\
\bottomrule
\end{tabular}
}
\begin{minipage}{0.95\linewidth}\vspace{2pt}\footnotesize\emph{Notes:} Each row compares the two production batches after the event with the two before, with reviewer fixed effects (post minus pre). Standard errors in parentheses are heteroskedasticity-robust and treat recordings within a batch as independent. The model and code are the same for every batch between two deploys, but Section~\ref{sec:drift} shows the agent's state also moved between batches without a deploy, and whatever moves it (an unlogged change, a vendor update behind a fixed version name, a shift in who was interviewed that fortnight) hits every recording in a batch together; each contrast therefore rests on four batch-level draws, the standard errors understate the uncertainty, and no test with four batches is reliable. We mark no significance and read the contrasts as magnitudes against each other: large in April and late July, close to zero in mid-July and at the vendor launch. The production models ran under fixed version names, so the 23 April vendor launch changed nothing in production and serves as a placebo date.\end{minipage}
\end{table}

%% file: sections/05_drift.tex
\section{How long does the evidence stay useful?}\label{sec:drift}

An evaluation result is a reading of a state that moves. How fast it moves, and how much of the movement the firm's own records account for, decides how long a reading remains informative.

\subsection{Specification}

The local-level model of Section~\ref{sec:method} returns, as a by-product of smoothing, the innovation variance $q$: the variance of the state's unpredictable movement over one week, so that its standard deviation over $h$ weeks grows as $\sqrt{qh}$. Section~\ref{sec:framework} distinguished total movement, with $q$ estimated on every transition between batches, from unaccounted movement, with the state free to jump across the three transitions that straddle a production deploy (13 April, 27 July and 10 August) so that $q$ is identified only from batches between which nothing was deployed. We estimate both, and a third variant that also frees the two late-May transitions carrying the step that Section~\ref{sec:deploys} could not attribute to any deploy. A likelihood-ratio test of $q=0$ asks whether the remaining transitions are consistent with a constant state. Forecast uncertainty $h$ weeks after the last batch has variance $P_T + qh$, where $P_T$ is the smoothed variance at the last batch.

\subsection{How fast the state moves, and how much of it the change log explains}

Table~\ref{tab:driftsources} reports the three estimates. For the index, the standard deviation of a week's total movement is \sdIndexWeek{} index standard deviations (95\% likelihood-ratio interval \sdIndexWeekLo{} to \sdIndexWeekHi) and of a week's unaccounted movement \sdIndexWeekFree{} (\sdIndexWeekFreeLo{} to \sdIndexWeekFreeHi); both reject a constant state at $p\qIndexFreeP$. Taking the deploys out changes the estimate hardly at all: most of the index's measured movement happened between batches with no logged production deploy. That movement is not diffuse either. Once the late-May step is also freed, the remaining transitions are consistent with a constant state (estimate \sdIndexWeekNoStep, upper bound \sdIndexWeekNoStepHi, $p=\qIndexNoStepP$). The index moved in a few steps, one at the April deploy, one in late May with nothing in the change log behind it, and one across the late-July deploys, rather than continuously. For the critical-failure rate, a week's total movement has a standard deviation of \sdCritWeek{} percentage points and the unaccounted movement \sdCritWeekFree{} ($p=\qCritFreeP$); freeing the late-May step reduces it to \sdCritWeekNoStep{} ($p=\qCritNoStepP$), so critical failures moved more gradually and part of their drift survives both exclusions, though not significantly.

\input{tables/tab_drift_sources}

Three sources could produce the unaccounted movement, and the data cannot separate them. A change may have reached production without a commit on the production branch (configuration, infrastructure, or prompt tables kept outside the repository), which the engineering team could check. The vendor may have updated the model served under a fixed version name (Section~\ref{sec:setting}). Or the interviews may have changed: the job records show the share of software roles roughly doubling between mid-March and early May and the acquisition channel shifting from June, and neither dimension is observed for the evaluated recordings themselves. We therefore describe the estimate as movement the change log does not account for, from whatever source, and not as drift in the agent alone. For the firm the distinction matters less than it does for us: the movement happened, it was as large between deploys as across them, and the firm's own records would not have shown it.

\subsection{From drift to a refresh interval}

\begin{figure}[t]
\centering
\includegraphics[width=\textwidth]{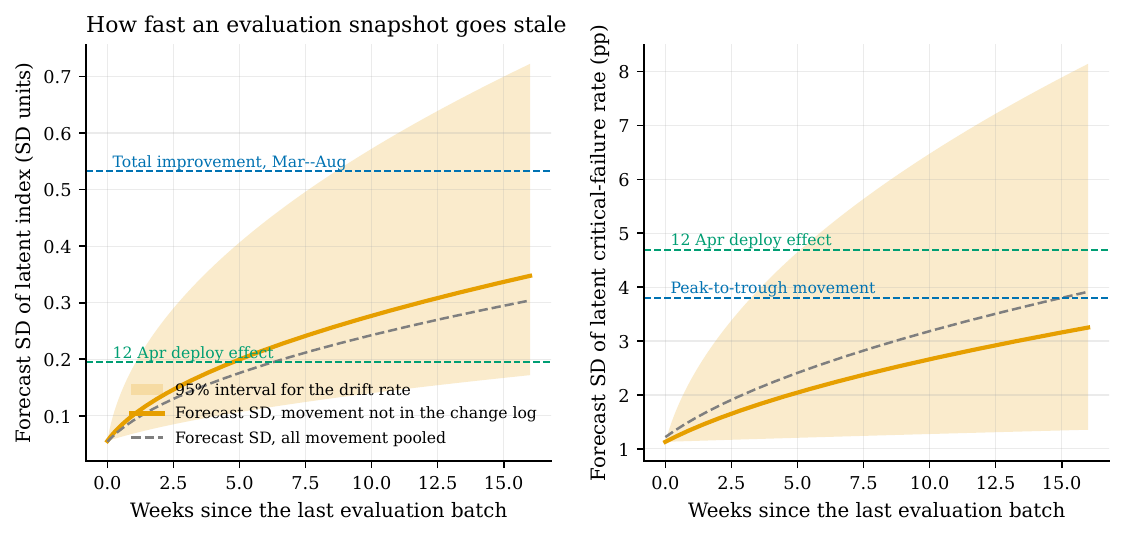}
\caption{Forecast standard deviation of the latent state as a function of weeks since the last evaluation batch, from the unaccounted-movement rate (solid, with the 95\% interval for the rate shaded) and from the pooled rate (dashed). Reference lines show the 12 April deploy contrast and the total movement observed over the window.}
\label{fig:refresh}
\end{figure}

The reason to want the drift rate in these units is Proposition~2 of Section~\ref{sec:framework}: the optimal interval between evaluations depends on the drift rate, the cost of a batch and the loss weight, and the horizon at which a reading stops being informative depends on $q$ and the uncertainty at the last reading. We do not observe the cost of a batch or the loss weight, so we do not compute an optimal interval; we report the horizon, which needs only $q$, $P_T$ and the size of a change the firm cares about. One condition attaches to it: the calculation treats the unaccounted movement as a diffusion, whereas Table~\ref{tab:driftsources} showed that for the index it was concentrated in one step, so the horizon is best read as the point at which a firm relying on its records would have been wrong by a deploy-sized amount, not as a smooth forecast. Suppose the firm stops evaluating after the last batch and deploys nothing. Its uncertainty about the agent's state $h$ weeks later has variance $P_T + q h$, and the forecast standard deviation grows with the square root of elapsed time. Figure~\ref{fig:refresh} plots it, with the pooled rate as a dashed comparison. For the index, the forecast standard deviation is \refreshIndexFour{} standard deviations after four weeks, \refreshIndexEight{} after eight and \refreshIndexTwelve{} after twelve. It reaches the size of the largest observed deploy contrast, the 12 April change of \aprIndex{} standard deviations, after about \refreshIndexWeeks{} weeks at the point estimate of the drift rate (\refreshIndexWeeksAll{} with the pooled rate), and between \refreshIndexWeeksLo{} and \refreshIndexWeeksHi{} weeks across the 95\% interval for the rate. The interval is wide because fourteen batches identify a variance loosely; the point estimate says that after about five weeks without evaluation and without a deploy, the firm could no longer tell whether a change as large as its biggest deploy had happened. For the critical-failure rate the forecast standard deviation is \refreshCritFour{} percentage points at four weeks, \refreshCritEight{} at eight and \refreshCritTwelve{} at twelve, against a peak-to-trough movement of about four points over the window; a quarter without evaluation would leave the firm unable to say whether it was at the top or the bottom of the range it had experienced.

Two things follow for the firm. First, at its fortnightly cadence the forecast standard deviation is about half the size of a large deploy contrast, so a batch of 300 recordings could still detect such a change between two readings; whether a fortnight was the right interval also depends on the cost of a batch and the loss from unreliability, which we do not observe. Second, the rate to use is the movement the firm's own change log does not account for, since deploys are known in advance and can be followed by an evaluation batch as a matter of routine. The number is specific to this agent and this period: an agent whose configuration is fully version-controlled and whose vendor serves dated snapshots would show less unaccounted movement, and an agent with a floating model alias or a changing candidate population would show more. The local-level model is a few lines of code and gives each firm its own number rather than a slogan; it returns the parameter the control-chart literature has asked for since \citet{duncan1956}.

%% file: tables/tab_drift_sources.tex
\begin{table}[t]
\centering
\footnotesize
\caption{Where the drift comes from: total movement versus movement the change log does not account for}
\label{tab:driftsources}
\resizebox{\ifdim\width>\linewidth\linewidth\else\width\fi}{!}{%
\begin{tabular}{lrrr}
\toprule
Outcome & Transitions the state may jump across freely & Innovation SD, one week [95\% CI] & LR p, no drift \\
\midrule
Stability index & all transitions & 0.075 [0.041, 0.142] & 0.000 \\
 & state free to jump at the three deploys & 0.086 [0.041, 0.180] & 0.000 \\
 & also free across the late-May step (9 May to 1 June) & 0.000 [0.000, 0.151] & 0.500 \\
Critical failure & all transitions & 0.930 [0.263, 2.112] & 0.002 \\
 & state free to jump at the three deploys & 0.762 [0.186, 2.016] & 0.010 \\
 & also free across the late-May step (9 May to 1 June) & 0.666 [0.000, 2.392] & 0.163 \\
\bottomrule
\end{tabular}
}
\begin{minipage}{0.95\linewidth}\vspace{2pt}\footnotesize\emph{Notes:} Each row re-estimates the local-level drift rate after allowing the latent state to jump without penalty across the listed transitions, so that the drift rate is identified only from the remaining ones. Freeing the three deploy transitions leaves the estimate for the index essentially unchanged: most of its measured movement occurred between batches with no logged production deploy. Freeing the late-May step as well removes it: the index moved in a few steps rather than continuously. Critical failures moved more gradually and part of their drift survives both exclusions.\end{minipage}
\end{table}

%% file: sections/06_decisions.tex
\section{When does evaluation improve the firm's decisions?}\label{sec:decisions}

Proposition~1 of Section~\ref{sec:framework} says that evaluation has value only through a repair that depends on it, and Proposition~3 says the value is decided by what repair capacity lets the firm do with a finding. Neither can be estimated here, since no part of the deployment went unevaluated; what the record can show is whether findings were available when decisions were made, whether the pre-release channel carried a usable reading, and whether a market signal moved the way the evaluation did. This section reports those three things and then asks what the program would have to be worth to cover its cost.

\subsection{Findings arrived every fortnight; production changed three times}\label{sec:cadence}

\begin{figure}[t]
\centering
\includegraphics[width=0.8\textwidth]{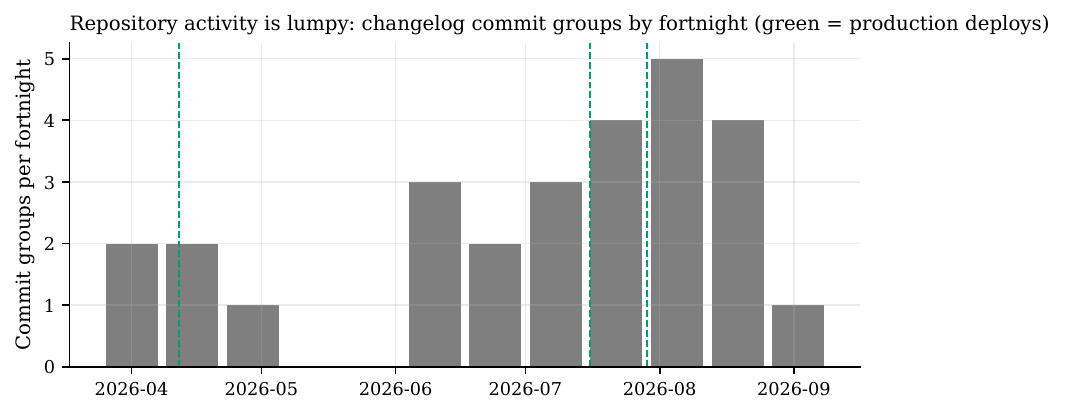}
\caption{Changelog commit groups per fortnight, 26 March to 31 August 2026. Green dashed lines mark the three production deploys.}
\label{fig:cadence}
\end{figure}

Figure~\ref{fig:cadence} plots commit groups per fortnight: \cadMin{} in two fortnights of April and May, \cadMax{} at the peak in July and August, and \nProdDeploys{} production deploys in five months. Commit groups measure activity in the repository rather than capacity or backlog, so the figure cannot show that engineers were constrained; what it shows is that findings arrived every fortnight throughout while production changes arrived three times, so that the interval between a finding and a change was long relative to the evaluation interval. That pattern points to repair, rather than information, as the scarcer input at the margin during the window, though a release policy that bundles changes into few deploys would leave the same trace. The spring, where critical failures rose for six weeks after the April deploy before falling, reads as a finding waiting for a repair.

\subsection{The pre-release channel as a test of remedies}\label{sec:prerelease}

\begin{figure}[t]
\centering
\includegraphics[width=0.85\textwidth]{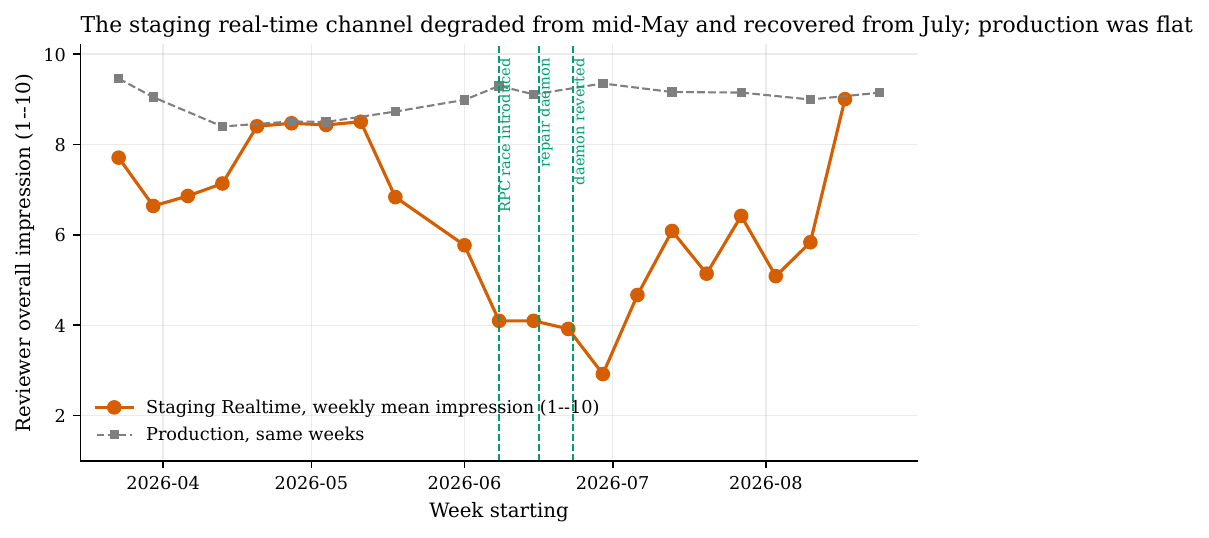}
\caption{Weekly mean reviewer impression (1 to 10) on the staging real-time channel and in production, with the three June development commits on reconnection. The staging channel was scored almost entirely by one reviewer, so the movement is within-reviewer.}
\label{fig:staging}
\end{figure}

The staging real-time channel, scored by the same pipeline and almost entirely by one reviewer, is where a finding that stops a regression before it reaches candidates would leave its trace (Figure~\ref{fig:staging}). Weekly mean impressions on staging ran at \stgMayEleven{} in the week of 11 May, fell to \stgMayEighteen{} the following week and \stgJunOne{} by the start of June, to \stgJunEight{} in the week of 8 June, and bottomed at \stgJunTwentyNine{} in the week of 29 June; they recovered to \stgJulThirteen{} by mid-July, ran between 5 and 6.5 through early August, and stood at \stgAugSeventeen{} in the final week, on \stgAugSeventeenN{} recordings. Production impressions over the same weeks are flat. The changelog shows a run of reconnection work on the development branch in June: an 8 June commit added a reconnection call and then delayed it, introducing a race; a 16 June commit added a repair daemon for stuck sessions with a ten-second startup cost; a 23 June commit reverted the daemon and added the question tracker.

The timing fits staging carrying experimental reconnection work while production did not, with the pipeline recording the degradation as it happened, though it does not line up commit by commit: the decline began in mid-May, three weeks before the first of these commits, and the trough came in the week after the revert. Whether the revert was prompted by the pipeline's reading of staging or by the engineers' own testing is not recorded, and no commit in the changelog cites an evaluation finding. Two things the episode does show. The pipeline saw a regression on the pre-release channel weeks before any production deploy, which is the information a test of remedies needs, and the regression never reached production. And the regression registered in the trainer's overall judgment rather than in the technical checklist: on the \stgNBefore{} staging recordings from the four weeks before the decline and the \stgNTrough{} at the bottom, the mean impression fell from \stgImprBefore{} to \stgImprTrough{} while the nine family scores barely moved (coherence from \stgCohBefore{} to \stgCohTrough{}, every other family at or above \stgOtherFamMinTrough{} in the trough) and the index was flat (\stgIdxBefore{} to \stgIdxTrough). Whatever the reviewer saw lay outside the rubric's sub-items, which is a reason to keep a holistic item on a technical form.

\subsection{Specification}

The ticket dashboard records daily counts by category. The ticket outcome is the log ratio of interview-related tickets to technical-issue tickets on the same dashboard, regressed on time in months over windows that start on 5 May (the full window), in June and in May and end in August or September; the same regressions are run on log interview tickets per interview, using the imputed volume series of Section~\ref{sec:setting}. A two-period comparison splits the window at each of five dates from 8 June to 15 July and compares the post-minus-pre change in the ticket ratio with the post-minus-pre change in three evaluation outcomes fixed in advance (the reviewer-adjusted index, the critical-failure rate and the share of interviews with any deduction). Both are descriptive: they establish direction and rough size, and the qualifications that attach to each are stated beside the result.

\subsection{Tickets against a same-instrument control}\label{sec:outcomes}

From 26 March to 4 May 2026 the firm had no consistently recorded market-facing measure of the interviewing agent's reliability. In those six weeks the pipeline recorded the rise in critical failures after the 12 April deploy, the fall in structure and audio-visual sync deductions, and the beginning of the latency improvement; whatever the firm knew about the agent's reliability that survives in a record, it knew from the pipeline, and whether the pipeline was reading the agent correctly can be asked only once a second instrument exists.

\begin{figure}[t]
\centering
\includegraphics[width=0.9\textwidth]{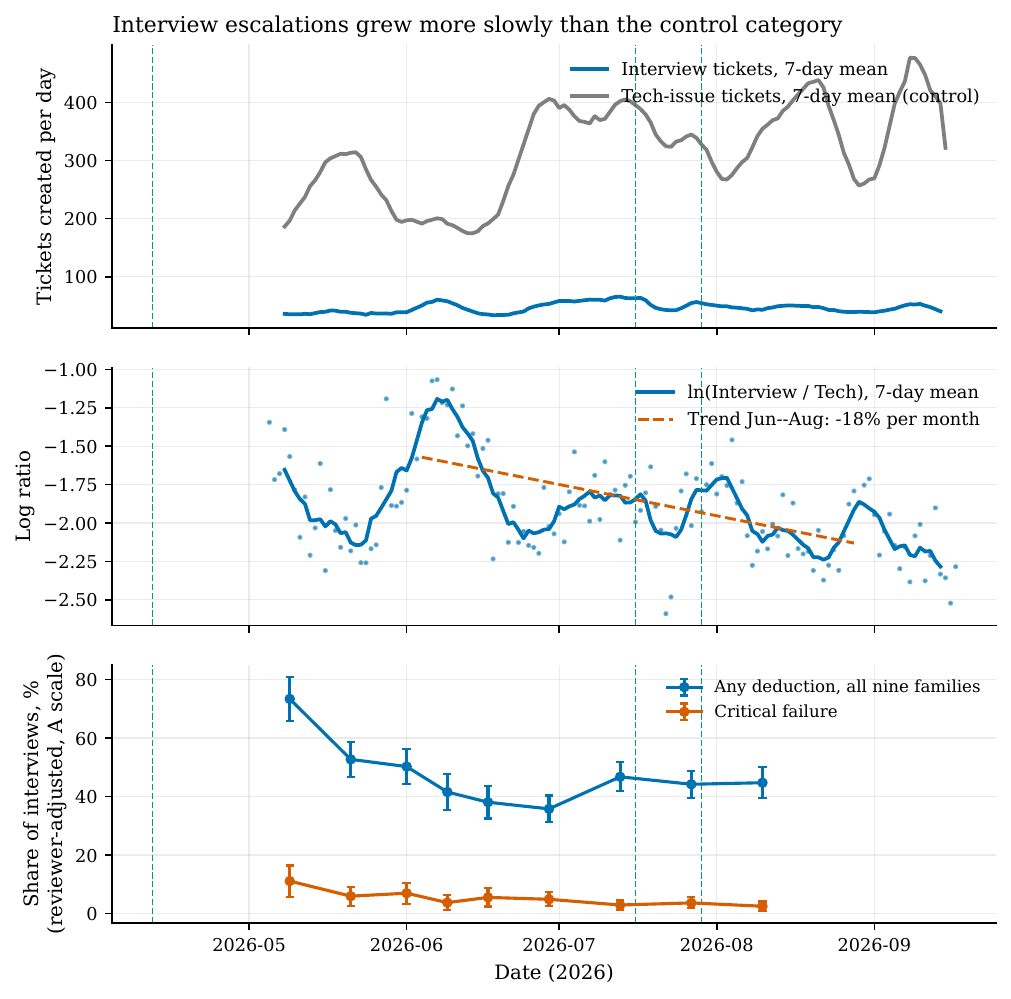}
\caption{Top: daily interview-related and technical-issue tickets created, seven-day centered means. Middle: log ratio of the two, with the June to August trend. Bottom: reviewer-adjusted share of production interviews with a deduction in any of the nine families, and the critical-failure rate, over the same weeks. Green dashed lines mark production deploys.}
\label{fig:tickets}
\end{figure}

The ticket dashboard launched on 4 May. Figure~\ref{fig:tickets} shows daily interview-related and technical-issue tickets as seven-day means, the log ratio of the two, and the reviewer-adjusted deduction share and critical-failure rate over the same period. Interview tickets were roughly flat in level over the full window from 5 May to 17 September (\tickLevelFull\% per month, $t=\tickLevelFullT$) while technical-issue tickets on the same dashboard grew, so the ratio fell: by \tickFullAbs\% per month over the full window ($t=\tickFullT$), the headline figure, and by \tickJunAugAbs\% over June to August ($t=\tickJunAugT$), a window that starts near a local peak of the ratio and is reported because it is where the two instruments overlap most fully (Table~\ref{tab:tickets}). The second column of Table~\ref{tab:tickets} controls instead for how many interviews ran: tickets per interview fell \tickVolFullAbs\% per month over the full window ($t=\tickVolFullT$) and \tickVolJunAugAbs\% from June to August ($t=\tickVolJunAugT$), while interview volume roughly doubled between May and September. The two denominators tell the same story. The qualification is that a fall in filed complaints is a fall in filed complaints: only some candidates who experience a failure file one, and the acquisition channel shifted over the summer toward candidates with no colleague at the firm, who may be readier to complain, which would push tickets per interview up, against the decline observed. We read the ticket series as a direction and an order of magnitude, not as an elasticity.

\input{tables/tab_tickets}

Table~\ref{tab:twoperiod} asks a simpler question: over the period both instruments cover, do they move the same way at every split? The index rises by \twoIndexMin{} to \twoIndexMax{} standard deviations depending on the split, the critical-failure rate falls by \twoCritMaxAbs{} to \twoCritMinAbs{} points, the any-deduction share by \twoAnyMaxAbs{} to \twoAnyMinAbs{} points, and the ticket ratio by \twoTicketMaxAbs\% to \twoTicketMinAbs\%. Both series are close to monotone over these weeks, so agreement at every split establishes shared direction and nothing more; a common trend in something else would produce the same table. What the agreement establishes is that the reading the pipeline gave the firm in the spring, that the agent's technical behavior was improving, is the reading the ticket record gives in retrospect, and that the raw averages, which flattened in June, would not have given it. It is a corroboration rather than a validation: the smoothed path uses batches on both sides of each date, and the decline is relative, since interview tickets did not fall in level but failed to grow while volume and the control category did.

\input{tables/tab_two_period}

\subsection{What the program would need to be worth}\label{sec:attribution}

A firm still has to decide whether to fund the program, and the record does not contain the number the decision would ideally use. The program's effect on the agent's reliability is not identified: no part of the deployment went unevaluated, so there is no comparison arm, and nothing varied the evaluation's intensity. The two designs one would reach for do not supply it. A difference-in-differences against the staging channel around the 16 July deploy gives \didIndex{} standard deviations on the index, but staging receives changes before production by construction, and the estimate is in any case an effect of a deploy rather than of the program. Commits do not cluster after sign-off dates (\timeObsSeven\% of the \nCommits{} commit groups fall within seven days of one against a permutation null of \timeNullSeven\%, $p=\timePSeven$), and a simulation in which every commit follows a finding at lags spread over three weeks rejects in only \powerTwentyOne\% of replications, so the timing evidence neither supports nor rules out the channel. What would identify the effect is a record, kept at the time, of which finding motivated each commit, and designed variation in evaluation intensity over time; staggering the introduction of evaluation across segments would not, because a repair to a shared agent reaches every segment at once. Both have to be designed in before the program starts.

Without the causal number, the decision can still be posed as a condition. On the cost side, the stability pipeline's rater pay was \$\costMonthly{} per month averaged over six months, from payroll; engineering time spent acting on findings is not counted. On the benefit side we count one channel only, the interview-related escalations that candidates did not file, with the caveat already stated that fewer filed complaints need not mean fewer failures \citep{fornell1987}. Relative to the first four weeks of the dashboard, interview tickets in September ran at about \actualMonthlyEnd{} per month against about \referenceMonthlyEnd{} had they kept their May ratio to technical-issue tickets: a reduction of \avoidedEndTech{} per month, or \reductionShareEnd\%. Averaged over June to September the reduction is much smaller, because tickets per interview rose in June and July before falling, and a June baseline would put it near \avoidedJuneBase; we use the May baseline and the September level.

Table~\ref{tab:breakeven} puts the two together. At the September reduction and full credit, the program cost \$\costPerAvoidedTech{} per avoided escalation. If handling one escalation costs the firm \$100, including the support time and the candidate's experience, the program covers its cost on this channel if it deserves credit for about \creditHundredTech\% of the September reduction; at \$50 it needs credit for about twice that share, and at \$25 for most of it. The point is not the return, which the table gives under full credit, but the form of the question: the funding decision reduces to whether the evaluation program deserves a fifth of one channel's improvement, which is a question about mechanism that engineering can judge, rather than a return on investment, which the data cannot deliver.

\input{tables/tab_breakeven}

%% file: tables/tab_tickets.tex
\begin{table}[t]
\centering
\footnotesize
\caption{Interview-related support tickets: monthly trend against same-instrument controls}
\label{tab:tickets}
\resizebox{\ifdim\width>\linewidth\linewidth\else\width\fi}{!}{%
\begin{tabular}{lrrrr}
\toprule
Window & ln(Interview) - ln(Tech issues) & ln(Interview) - ln(interview volume) & ln(Interview tickets), no control & ln(Interview) - ln(other categories) \\
\midrule
May 5 -- Sep 17 (full) & -9.9\% (t = -2.69) & -7.2\% (t = -2.20) & +3.7\% (t = +1.43) & -23.2\% (t = -3.74) \\
May 5 -- Aug 31 & -8.0\% (t = -1.77) & -3.8\% (t = -0.93) & +6.3\% (t = +1.99) & -15.0\% (t = -2.80) \\
Jun 1 -- Aug 31 & -18.1\% (t = -2.85) & -14.1\% (t = -2.68) & -1.8\% (t = -0.35) & -26.8\% (t = -4.61) \\
Jun 1 -- Sep 17 & -17.7\% (t = -3.84) & -16.0\% (t = -4.03) & -3.0\% (t = -0.74) & -34.1\% (t = -5.44) \\
\bottomrule
\end{tabular}
}
\begin{minipage}{0.95\linewidth}\vspace{2pt}\footnotesize\emph{Notes:} Daily ticket creation from the support dashboard (launched 4 May 2026; the part day is dropped). Each cell is the implied percentage change per month from a regression of the log series on a linear time trend and day-of-week dummies, with Newey-West standard errors (7 lags). The Tech-issues category is the primary control because it shares the instrument and the user base; the interview-volume column divides by the approximate daily number of AI interviews from the job records (Section 2), so it reads as tickets per interview (the job records end on 15 September, so that column stops there); 'other categories' is the residual of all types and is inflated by category rollout, so its column overstates the decline.\end{minipage}
\end{table}

%% file: tables/tab_two_period.tex
\begin{table}[t]
\centering
\footnotesize
\caption{Two instruments at alternative break dates within the ticket window: post-minus-pre changes}
\label{tab:twoperiod}
\resizebox{\ifdim\width>\linewidth\linewidth\else\width\fi}{!}{%
\begin{tabular}{lrrrr}
\toprule
Break date & Stability index, SD & Critical failure, pp & Any deduction, pp & Interview/Tech ticket ratio \\
\midrule
2026-06-08 & +0.18 (0.09)* & -3.76 (1.28)** & -14.76 (5.68)** & -10.3\% (t = -0.91) \\
2026-06-15 & +0.14 (0.07)* & -2.74 (1.27)* & -10.67 (5.42)* & -21.5\% (t = -1.79) \\
2026-06-25 & +0.10 (0.07) & -2.80 (1.03)** & -6.67 (5.15) & -19.9\% (t = -1.89) \\
2026-07-05 & +0.05 (0.06) & -2.89 (0.79)*** & -0.91 (4.23) & -16.4\% (t = -1.60) \\
2026-07-15 & +0.06 (0.06) & -2.23 (0.87)** & -1.82 (3.41) & -17.3\% (t = -1.83) \\
\bottomrule
\end{tabular}
}
\begin{minipage}{0.95\linewidth}\vspace{2pt}\footnotesize\emph{Notes:} Eval columns: change in the reviewer-adjusted composite index, in the critical-failure rate, and in the share of production interviews with a deduction in any of the nine families, production batches from 4 May to 10 August 2026, reviewer fixed effects, standard errors clustered on batch day; the outcomes are fixed in advance and not selected on their movement. Ticket column: percentage change in the daily Interview/Tech-issues ratio (day-of-week controls, Newey-West). Both series are close to monotone over these weeks, so agreement at every split shows shared direction and cannot separate a shared cause from parallel ones.\end{minipage}
\end{table}

%% file: tables/tab_breakeven.tex
\begin{table}[t]
\centering
\footnotesize
\caption{Share of the escalation reduction the agent evaluation program must account for to cover its cost}
\label{tab:breakeven}
\resizebox{\ifdim\width>\linewidth\linewidth\else\width\fi}{!}{%
\begin{tabular}{lrr}
\toprule
Cost of one escalation & Credit needed & Net value per month at full credit \\
\midrule
\$25 & 73\% & \$2,905 \\
\$50 & 36\% & \$13,616 \\
\textbf{\$100} & \textbf{18\%} & \textbf{\$35,037} \\
\$150 & 12\% & \$56,459 \\
\$250 & 7\% & \$99,303 \\
\bottomrule
\end{tabular}
}
\begin{minipage}{0.95\linewidth}\vspace{2pt}\footnotesize\emph{Notes:} Program cost is the rater pay for the stability evaluation pipeline, \$7,806 per month averaged over six months; engineering time spent acting on findings is not included. The escalation reduction is how many fewer interview tickets per month the firm received in September 2026 than it would have if interview tickets had kept the ratio to technical-issue tickets they had in the first four weeks of the dashboard: 428 per month. The cost of one escalation is not observed, so the rows span a range; \$100, in bold, is the value used in the text. ``Credit needed'' is the share of the reduction the evaluation program would have to be responsible for to cover its cost at that cost per escalation. ``Net value'' is the saving at that cost minus the program cost, if the program is given credit for the whole reduction.\end{minipage}
\end{table}

%% file: sections/07_conclusion.tex
\section{Conclusion}\label{sec:conclusion}

Evaluating a deployed AI agent is a problem of measuring change while both the system and its measurement process evolve. In this deployment, reviewer composition concealed meaningful improvement in the evaluation index. Adjusting for that composition revealed a clearer performance path, while family-level analysis showed that gains in structured behavior and changes in severe failures followed different trajectories. Performance also changed outside recorded production releases, demonstrating why release histories alone cannot establish that earlier evaluation results remain informative.

The support-ticket record provides a complementary perspective: interview-related support burden declined relative to both another ticket category and estimated interview volume. This agreement connects the evaluation results to operational experience without treating improvements observed during an evaluation program as effects caused by that program.

Together, the findings support an approach to evaluation centered on comparable measurement, repeated observation, and decision-relevant outcomes. The purpose is not simply to produce more scores, but to provide evidence that helps firms distinguish progress from changes in measurement, identify consequential weaknesses, and decide when their existing evidence needs to be refreshed.

%% file: sections/appendix.tex
\section{Proofs for Section~\ref{sec:framework}}\label{app:model}

\paragraph{Posterior dynamics.} With the state equation \eqref{eq:state} and the batch reading \eqref{eq:batchmean} net of known rater effects, the firm's posterior is Gaussian with mean $\hat\theta_t$ and variance $P_t$. Between batches the mean advances by the known repair effect $g(e_t)$ (the drift has mean zero) and the variance grows linearly, $P_{t+s} = P_t + q\,s$. At a batch with sampling variance $R = \sigma^2/n$ the variance updates to $P^{+} = \big(1/P^{-} + 1/R\big)^{-1}$, where $P^{-}$ is the variance just before the batch; $P^{+} \le R$ and $P^{+} \le P^{-}$, and $P^{+}$ approaches $R$ from below as $P^{-}$ grows. These are the Kalman filter recursions for the local-level model \citep{harvey1989,durbin2012}. In steady state with a fixed interval $h$, $P^{-} = P^{+} + qh$, which gives the quadratic $P^{+\,2} + qh\,P^{+} - qhR = 0$ and hence
\[
P_0(h,n) \equiv P^{+} = \frac{\sqrt{(qh)^2 + 4qhR} - qh}{2}.
\]

\paragraph{Proof of Proposition 1.} Write the firm's expected loss from unreliability under an evaluation policy $\mathcal{E}$ and a repair rule $e(\cdot)$ as $\mathcal{L}(\mathcal{E}, e)$. If $e$ does not depend on the information $\mathcal{E}$ generates, then $\mathcal{L}(\mathcal{E}, e) = \mathcal{L}(\varnothing, e)$ for every $\mathcal{E}$, since the distribution of $\theta_t$ is the same under both and the loss depends only on $\theta_t$ and $e_t$; the gross value of evaluation is zero and its net value is $-\sum_t c(n)\mathbf{1}\{\text{batch in }t\}$. With rules optimal for their information, the gross value $\mathcal{L}(\varnothing, e^{\varnothing}) - \mathcal{L}(\mathcal{E}, e^{\mathcal{E}})$ is non-negative \citep{blackwell1953} and equals zero when the optimal rule is the same under both, which holds when $\bar e = 0$. In the general problem the gross value is the reduction in minimized expected loss, repair cost included, and has no closed form. Under the quadratic monitoring approximation of Section~\ref{sec:framework} (quadratic decision loss, linear and unconstrained repair) the gross value of a batch that moves the posterior variance from $P^{-}$ to $P^{+}$ and is acted on is $\kappa\,(P^{-} - P^{+})$ per week until the next batch.

\paragraph{Proof of Proposition 2.} Under the same approximation, over an interval of length $h$ the firm acts on an estimate whose error variance $s$ weeks after the last batch is $P_0 + q\,s$; the expected excess loss is $\kappa\,(P_0 + q s)$ and its average over the interval is $\kappa\,(P_0 + q h/2)$, with $P_0 = P_0(h,n)$ from the steady-state expression above. Adding the evaluation cost per week, $c(n)/h$, gives the objective in \eqref{eq:interval}. Because $\partial P_0/\partial h > 0$ and $\partial P_0/\partial q > 0$, the marginal cost of a longer interval exceeds $\kappa q/2$ and the optimum $h^*$ satisfies $c(n)/h^{*2} = \kappa\,(q/2 + \partial P_0/\partial h)$; by the implicit function theorem $\partial h^*/\partial q < 0$, $\partial h^*/\partial \kappa < 0$ and $\partial h^*/\partial c > 0$. When $R = \sigma^2/n$ is small relative to $qh$, $P_0 \approx R - R^2/(qh) \to R$, its derivative in $h$ vanishes, and the first-order condition reduces to $c(n)/h^2 = \kappa q/2$, hence $h^{*} = \sqrt{2c(n)/(\kappa q)}$. The horizon at which the forecast variance reaches $D^2$ solves $P_0 + q s = D^2$, giving $s = (D^2 - P_0)/q$; in the text $P_0$ is the smoothed variance at the last batch rather than the steady-state value.

\paragraph{Proof of Proposition 3.} Let two failure modes have shortfalls $\delta_1, \delta_2 \ge 0$ below their targets, unknown to the firm before evaluation, with a common prior, and let one unit of repair effort remove one shortfall entirely. The value of information is zero exactly when the optimal repair is the same under every signal \citep{raiffa1961,blackwell1953}. With $\bar e = 0$ nothing is repaired under any information, so the value is zero. With $\bar e \ge 2$ and costless repair ($w=0$) both modes are repaired under any information, so the value is again zero. With $\bar e = 1$ the firm repairs the mode it believes worse; without information it picks at random and expects to remove $\mathbb{E}[\delta]$; with perfect information it removes $\mathbb{E}[\max(\delta_1, \delta_2)]$; the value of evaluation is $\kappa$ times the difference, which is strictly positive whenever the shortfalls differ with positive probability. With costless repair the value is thus zero at $\bar e = 0$, positive at $\bar e = 1$ and zero at $\bar e = 2$: non-monotone in capacity. With costly repair the high-capacity value is positive rather than zero, because the finding tells the firm which repairs are worth their cost, but it stays below the value under the binding constraint. Take two modes of which exactly one, with equal probability, carries a loss of 10, and a repair cost of 1: an uninformed firm with capacity for two repairs makes both for an expected net gain of 8, an informed firm makes only the useful one for 9, so the information is worth 1; with capacity for one repair the uninformed firm expects 4 and the informed firm 9, so it is worth 5. In either case evaluation and capacity are neither complements nor substitutes in general, and the marginal value of evaluation under a binding constraint is zero only if the firm already knows which mode is worse. The observable implication in the text (findings that wait for capacity indicate that repair is the scarcer input) follows from the queue, not from a sign on a cross-partial.

\paragraph{Proof of Proposition 4.} Equation \eqref{eq:bias} follows from differencing \eqref{eq:batchmean}. Suppose sessions are assigned to raters within a batch independently of $\varepsilon_{irt}$ (exchangeable assignment) and $\gamma_r$ is constant over time. Then for two raters $r, r'$ who both scored batch $t$, $\mathbb{E}[y_{irt} - y_{ir't}] = \gamma_r - \gamma_{r'}$; for a rater $r$ who scored batches $t$ and $t'$, $\mathbb{E}[y_{irt'} - y_{irt}] = \theta_{t'} - \theta_t$; and if the rater-by-batch graph is connected, both differences extend to every pair, so the parameters of \eqref{eq:meas} are identified up to one normalization \citep{abowd1999,abowd2002}. Setting $\gamma_{r_0} = 0$ for the rater who scored every batch fixes the level of $\theta_t$ on that rater's scale. Without exchangeable assignment the within-batch difference identifies $\gamma_r - \gamma_{r'}$ plus the difference in mean difficulty of the sessions each rater received; without constant severity a common drift in raters' standards is indistinguishable from movement in $\theta_t$. Section~\ref{sec:measurement} tests the second condition partially (the per-batch gap between the two production raters) and checks the first indirectly (the reviewer-A-only trend).

\section{Variable construction}\label{app:vars}

\paragraph{Stability pipeline.} Each of the nine families is scored as the share of its sub-items passed, on a 0 to 1 scale; a deduction indicator equals one if any sub-item was lost. The stability index is the unweighted mean of the nine family scores. Standardized versions (SD units) subtract the mean and divide by the standard deviation over all \nStabAll{} scored recordings, production and staging, so that production and staging can be compared on one scale. The critical-failure flag is one if the trainer selected any deal-breaker on the form; the broader ``any failure'' flag is one if the trainer either flagged a critical failure or gave an overall impression of 6 or below on the 1 to 10 scale. Batches are keyed to the interview date; the \nBatches{} production batch dates are 29 March, 5, 13 and 27 April, 9 and 21 May, 1, 9, 17 and 29 June, 13 and 27 July, and 10 and 24 August 2026. Deploy regimes split the window at the three production deploys (12 April, 16 July, 29 July).

\paragraph{Interview population.} From the job-posting records, each posting's AI interviews are spread uniformly over the days between its first and last recorded interview start and summed by day and by role family; role families are assigned by keyword match on the role title (software and engineering; generalist and annotation; professional and other as the residual). Applicants are attributed to the posting month of the job and grouped by the recorded acquisition source (referral, direct, search engines, paid advertising, community and partner channels, LinkedIn, internal invitations, and others). The records end on 15 September 2026, so the volume denominator is undefined after that date.

\paragraph{Tickets.} Daily counts of tickets created by category, 5 May to 17 September 2026. Log series use $\log(\max(x,1))$. Trend regressions include day-of-week dummies and use Newey-West standard errors with seven lags \citep{newey1987}.

\section{Estimation details}\label{app:est}

\paragraph{Fixed-effects regressions.} All regressions are ordinary least squares with dummy variables. Standard errors are clustered on batch day (\nBatches{} clusters in production) with the usual small-sample factor; because fourteen clusters is few, the main trend estimates are accompanied by a restricted wild cluster bootstrap $p$-value \citep{cameron2008} with the six-point weights of \citet{webb2023}, which \citet{mackinnon2018} recommend when clusters are few, and 999 draws. Contrasts around single dates rest on four batches and use heteroskedasticity-robust standard errors instead.

\paragraph{Stage-1 batch effects.} For each outcome the design contains all \nBatches{} batch dummies and dummies for trainers B and \rev{C}, with \rev{A} as the omitted reference; there is no intercept. The batch coefficients are the reviewer-adjusted batch effects on \rev{A}'s scale. Their sampling variances are heteroskedasticity-robust. For binary outcomes the variance is floored at $\tilde p(1-\tilde p)/n$ with $\tilde p=(k+0.5)/(n+1)$, so that a batch with zero events does not receive infinite weight in Stage 2.

\paragraph{Stage-2 local level.} The model is $\hat\theta_t=\mu_t+v_t$, $\mu_{t+1}=\mu_t+\eta_t$, with $(v_1,\dots,v_T)\sim N(0,\hat\Sigma)$, where $\hat\Sigma$ is the Stage-1 sampling covariance of the batch effects, and $\operatorname{Var}(\eta_t)=q\,\Delta_t$ for a gap of $\Delta_t$ weeks between batches. $\hat\Sigma$ is not diagonal: the batch effects share the estimated reviewer coefficients, so batches scored by two reviewers covary through $\hat\gamma_B$ (about $\operatorname{Var}(\hat\gamma_B)\,s_{Bt}\,s_{Bt'}$, roughly a tenth of a batch effect's sampling variance; the absolute correlations average \offdiagCorrIndex{} for the index). The likelihood is the exact diffuse likelihood of the first differences of the batch effects, $D\hat\theta\sim N(0,\,Q+D\hat\Sigma D')$ with $Q=\operatorname{diag}(q\,\Delta_t)$, which removes the initial level; with a diagonal $\hat\Sigma$ it coincides with the Kalman-filter likelihood initialized diffusely on the first batch. $q$ is estimated by maximizing the profile likelihood over a logarithmic grid followed by bounded refinement, with the constant-state model $q=0$ evaluated exactly; the 95\% interval collects the values whose likelihood lies within $\chi^2_1(0.95)/2$ of the maximum, and reports a zero lower bound when $q=0$ lies within that band; the test of $q=0$ uses the boundary-corrected likelihood ratio (half a $\chi^2_1$). Smoothed states and their covariance are the Gaussian posterior of $\mu$ given $\hat\theta$ under a flat prior on the initial level, which with a diagonal $\hat\Sigma$ reproduces the fixed-interval smoother \citep{durbin2012}; the standard error of the change between two batches uses the posterior covariance of the two endpoints. Transitions across which the state is allowed to jump freely (Table~\ref{tab:driftsources}) receive a large innovation variance, which removes their difference from the information about $q$. Forecast variances $h$ weeks past the last batch are $P_T+qh$. Carrying $\hat\Sigma$ rather than its diagonal alone changes the smoothed first-to-last change in the index from \latentIndexChangeDiag{} to \latentIndexChange{} standard deviations and the innovation SD from \sdIndexWeekDiag{} to \sdIndexWeek; the reviewer-A-only trend (Table~\ref{tab:trends}) involves no shared offset and is the independent check.

\paragraph{Break test.} The sup-$F$ statistic is computed on the Stage-1 batch effects with inverse-variance weights, for candidate breaks between the third and twelfth batch (15\% trimming), following \citet{andrews1993}. With fourteen estimated points the asymptotic critical values do not apply, and the statistic is used only to locate the largest single level shift.

\paragraph{Timing tests.} Commit groups are the \nCommits{} dated headings on the development and production branches of the changelog between 26 March and 31 August 2026; a heading that lists several dates counts once, at its first listed date. Sign-off events are, for each of the \nSignoff{} production batches, the date by which half of the batch's recordings had been signed off, taken from the sign-off timestamps in the export; the median review lags the interview date by about \signoffLag{} days, with the first sign-off one to three days after the interview date and the last several weeks later. The permutation null places \nCommits{} commits uniformly at random over the window and recomputes the share within $k$ days after a batch date, 4,000 draws. The phase test uses the position of each commit within the interval between the batch dates that bracket it; commits dated before the first or after the last sign-off date have no bracketing interval and are excluded (\phaseDropped{} of them), leaving \phaseN{} of \nCommits. The lag simulation draws a batch date and a lag for each commit and recomputes the seven-day share, \lagSimReps{} draws per lag distribution.

\section{Family-level trend regressions}\label{app:families}

Table~\ref{tab:famtrends} reports, for each rubric family, the trend in the standardized family score with reviewer fixed effects and with reviewer and deploy-regime fixed effects, the trend in the deduction indicator, and the Holm and Benjamini-Hochberg adjustments of the deduction-trend p-values across the nine families, which are the quantities summarized in the panel titles of Figure~\ref{fig:families}.

\input{tables/tab_family_trends}

%% file: tables/tab_family_trends.tex
\begin{table}[t]
\centering
\scriptsize
\caption{Where the improvement sits: trends by rubric family, production}
\label{tab:famtrends}
\resizebox{\ifdim\width>\linewidth\linewidth\else\width\fi}{!}{%
\begin{tabular}{lrrrrrrr}
\toprule
Family & Mean score & At ceiling & Trend, SD/month (reviewer FE) & Trend, SD/month (+ regime FE) & Deduction trend, pp/month & Holm p & BH q \\
\midrule
Conversation structure & 0.954 & 67\% & +0.111 (0.028)*** & +0.079 (0.042)* & -9.26 (1.62)*** & 0.001 & 0.001 \\
Interruption handling & 0.981 & 85\% & -0.012 (0.013) & +0.017 (0.028) & +0.21 (0.80) & 1.000 & 0.792 \\
Latency & 0.872 & 55\% & +0.055 (0.020)** & +0.083 (0.023)*** & -3.00 (1.26)** & 0.174 & 0.063 \\
Speech pacing & 0.996 & 98\% & +0.003 (0.011) & -0.000 (0.016) & -0.10 (0.07) & 0.504 & 0.216 \\
Coherence & 0.997 & 99\% & +0.003 (0.008) & +0.007 (0.006) & -0.22 (0.14) & 0.504 & 0.189 \\
AV sync & 0.983 & 96\% & +0.161 (0.040)*** & +0.197 (0.040)*** & -3.59 (1.34)** & 0.150 & 0.047 \\
Freezes & 0.997 & 99\% & +0.005 (0.020) & +0.036 (0.015)** & -0.06 (0.12) & 1.000 & 0.745 \\
Audio quality & 0.996 & 94\% & +0.016 (0.010) & +0.001 (0.021) & -2.06 (0.77)** & 0.150 & 0.047 \\
Interview outcome & 0.995 & 97\% & +0.015 (0.008)* & +0.027 (0.012)* & -0.83 (0.31)** & 0.150 & 0.047 \\
\bottomrule
\end{tabular}
}
\begin{minipage}{0.95\linewidth}\vspace{2pt}\footnotesize\emph{Notes:} Family scores are 0--1 (share of sub-items passed) standardized per family; 'deduction' is an indicator for any lost sub-item. Reviewer fixed effects; the fourth column adds deploy-regime fixed effects so the trend is identified within regimes only. Standard errors clustered on batch day (13 clusters); partial 24 August batch excluded. AV sync accounts for 73\% of critical failures. The last two columns adjust the deduction-trend p-values for the nine families tested (Holm step-down and Benjamini-Hochberg); the stars in the three trend columns are unadjusted. $^{*}p<0.10$, $^{**}p<0.05$, $^{***}p<0.01$.\end{minipage}
\end{table}